\documentclass[aps,12pt,prd,showpacs,notitlepage,nofootinbib,tightenlines]{revtex4-1}
\usepackage{amsmath}
\usepackage{bm}
\usepackage{times}
\usepackage{braket}
\usepackage{color}
\usepackage{epsfig}
\usepackage{slashed}
\usepackage{hyperref}
\newcommand{\beq}{\begin{eqnarray}}
\newcommand{\eeq}{\end{eqnarray}}
\newcommand{\non}{\nonumber\\ }

\newcommand{\etap}{\eta^{(\prime)} }

\newcommand{\mbs}{m_{B_s} }

\newcommand{\psl}{ P \hspace{-2.4truemm}/ }

\newcommand{\cala}{ {\cal A} }

\def \cpc{ Chin. Phys. C  }

\def \epjc{ Eur. Phys. J. C }
\def \jpg{  J. Phys. G }
\def \npb{  Nucl. Phys. B }
\def \plb{  Phys. Lett. B }
\def \ppnp{ Prog.Part. $\&$ Nucl. Phys. }
\def \prd{  Phys. Rev. D }
\def \prl{  Phys. Rev. Lett.  }

\def \jhep{ JHEP }
\def \rmp{ Rev. Mod. Phys. }
\definecolor{Red}{rgb}{1.,0.,0.}

\definecolor{Blue}{rgb}{0.,0.,1.}

\definecolor{nicered}{rgb}{0.7,0.1,0.1}
\definecolor{nicegreen}{rgb}{0.1,0.5,0.1}

\hypersetup{colorlinks,citecolor=nicegreen,linkcolor=nicered}
\begin{document}

\title{\boldmath Anatomy of  $B_s \to VV $ decays and effects of next-to-leading order contributions
in the perturbative QCD factorization approach}
\author{Da-Cheng Yan$^{1}$}  \email{1019453259@qq.com}
\author{Xin Liu$^{2}$}  \email{liuxin@jsnu.edu.cn}
\author{Zhen-Jun Xiao$^{1,3}$} \email{xiaozhenjun@njnu.edu.cn}
\affiliation{$^1$ Department of Physics and Institute of Theoretical Physics,
Nanjing Normal University, Nanjing, Jiangsu 210023, China}
\affiliation{$^2$ School of Physics and Electronic Engineering,
Jiangsu Normal University, Xuzhou 221116, China}
\affiliation{$^3$ Jiangsu Key Laboratory for Numerical Simulation of Large Scale Complex
Systems, Nanjing Normal University, Nanjing, Jiangsu 210023, China}
\date{\today}
\begin{abstract}
By employing the perturbative QCD (PQCD) factorization approach, we calculated
the branching ratios, CP-violating asymmetries, the longitudinal and transverse
polarization fractions and other  physical observables of the thirteen
charmless hadronic   $\bar{B}^0_s \to V V $ decays
with the inclusion of all currently known next-to-leading order (NLO) contributions.
We focused on the examination of the effects of all those currently known
NLO contributions and found  that:
(a) for the measured decays $\bar{ B}_s^0  \to \phi \phi, K^{*0} \phi,\bar K^{*0} K^{*0}$
and $\rho^0 \phi$, the NLO contributions can provide
$\sim 20\%$ to $\sim 40\%$ enhancements to the leading order (LO) PQCD predictions of
their CP-averaged branching ratios, and consequently the agreement
between the PQCD predictions and the measured values are improved
effectively after the inclusion of the NLO contributions;
(b) for the measured decays, the NLO corrections to the LO PQCD predictions for
$(f_L,f_\perp)$ and $(\phi_{\|},\phi_{\bot})$ are generally small in size,
but the weak penguin annihilation contributions play an important role
in understanding the data about their decay rates, $f_L$ and $f_\perp$;
(c) the NLO PQCD predictions for above mentioned physical observables do agree
with the measured ones and the theoretical predictions from the QCDF, SCET
and FAT approaches;
(d) for other considered $B_s^0 \to VV$ decays,  the NLO PQCD predictions for
their decay rates and other  physical observables are also
basically consistent with the theoretical predictions from other popular
approaches, future precision measurements could help us to
test or examine these predictions.
\end{abstract}


\vspace{1cm}

\maketitle

\newpage
\section{Introduction}

During the past three decades, the two-body charmless hadronic $B_s \to VV$ decays,
with $V$ being the light vector mesons $\rho, K^*, \phi$ and $\omega$,
have been studied by many authors based on rather different factorization approaches
\cite{qcdf2,qcdf07,qcdf09,qcdfwa,qcdfll,gf99,ali07,jpg06,pqcd2,scet,scetv}.
Several such decay modes, such as $B_s^0 \to \phi \phi$ decay,  have been observed by CDF and LHCb experiments
\cite{cdf,lhcbks,lhcbks1,lhcbphi,lhcbphi2,lhcbphiks,lhcbrhophi,lhcb0,pdg2016,hfag2016}.
When compared with the similar $B_s \to PP,PV$ (here $P= \pi,K,\eta$, and $\eta^\prime$)
decays,
$B_s \to VV$ are indeed much more complicated due to the fact that more helicity amplitudes should be
taken into account.
The $B_s \to VV$ decays can offer, consequently, rich opportunities for us to test the Stand Model (SM)
and to search for the exotic new physics beyond the SM.

Experimentally, a large transverse polarization fraction of $B \to \phi K^*$ was firstly observed
in 2003 by BABAR and Belle Collaborations \cite{babe}. The new world averages of $f_L$ as given by
HFAG-2016 \cite{hfag2016}  for $B\to (\phi,\rho,\omega)$ decays, for example, are the following:
\beq
f_L(B^+\to V K^{*+})&=& \left \{ \begin{array}{ll}
0.50\pm 0.05\;, & {\rm for} \ \ V=\phi,  \\
0.78\pm 0.12\;, & {\rm for} \ \ V=\rho^0,  \\
0.41\pm 0.19\;, & {\rm for} \ \ V=\omega,  \\ \end{array} \right. \label{eq:fl01a}, \\
f_L(B^0\to V K^{*0})&=& \left \{ \begin{array}{ll}
0.497\pm 0.017\;, & {\rm for} \ \ V=\phi,  \\
0.40\pm 0.14\;, & {\rm for} \ \ V=\rho^0,  \\
0.70\pm 0.13\;, & {\rm for} \ \ V=\omega,  \\ \end{array} \right. \label{eq:fl01b},
\eeq
These measured values were in strong confliction with the general expectation
$f_L \approx 1$ in the naive factorization ansatz~\cite{naive1}, which
is the so-called  ``polarization puzzle"~\cite{2004pft,2004pa,lipa,lipa1}.
The similar deviations also be observed later for $B^+ \to \phi (K_1,K_2^*)$ and $\omega K_2^*$ decays
\cite{Aubert:2008bc,hfag2016}.

For the charmless $B_s^0\to VV$ decays studied in this paper,
the similar puzzles have also been observed by CDF and LHCb Collaboration for
$\bar{B}_s^0 \to \phi\phi$, $ K^* \phi$ and $ K^{*0} \bar K^{*0}$
decay modes \cite{cdf,lhcbks,lhcbks1,lhcbphi,lhcbphi2,lhcbrhophi}. The new world averages of $f_L$ and $f_\perp$
as given by HFAG-2016 \cite{hfag2016}  for these three decay modes are the following:
\beq
\bar{B}_s^0 \to \phi \phi &:& \qquad  f_L=0.361\pm 0.022,  \quad  f_\perp =0.306\pm 0.023, \\
\bar{B}_s^0 \to K^* \phi &:& \qquad  f_L=0.51\pm 0.17, \quad  f_\perp =0.28\pm 0.12, \\
\bar{B}_s^0 \to K^{*0} \overline{K}^{*0} &:& \qquad  f_L=0.201\pm 0.070, \quad  f_\perp =0.38\pm 0.11.
\eeq
More measurements are expected in the near future.

Theoretically, a number of strategies were proposed to resolve the above mentioned
"polarization puzzle" within and/or beyond the SM.
For example, the weak penguin annihilation contributions in QCD factorization (QCDF) approach was proposed
by Kagan~\cite{Kann}, the final state interactions were considered in Refs.~\cite{final,fsi,2004pft},
the form-factor tuning in the perturbative QCD (PQCD) approach was suggested by Li~\cite{lipa},
and even the exotic new physics effects have been studied by authors in Refs.~\cite{np,np1}.
Obviously, it is hard to get a good answer to this seemingly long-standing puzzle at present.
However, according to the statement in Ref.~\cite{lipa1},  the complicated
QCD dynamics involved in such $B/B_s\to VV$ decays should be fully explored  before resorting to
the possible new physics beyond the SM.
Therefore, the QCDF approach ~\cite{qcdf2,qcdf07,qcdf09,qcdfwa,qcdfll,qcdfv},
the soft-collinear effective theory (SCET) ~\cite{scet,scetv} and
the PQCD approach ~\cite{ali07,Li2002,jpg06,pqcd2,pqcdv,pqcdv1}, have been adopted to
investigate these kinds of decays systematically.

The two-body charmless hadronic decays $B_s \to VV$ have been systematically studied in the PQCD
approach at leading order (LO) in 2007 \cite{ali07}.
Recently, the authoers of Ref.~\cite{pqcd2} made improved estimations for the $B_{(s)} \to VV$ modes
by keeping the terms with the higher power of the  ratios $r_{2,3} = m_{V_{2,3}}/m_{B_{(s)}}$
in the PQCD approach, with $m_{B_{(s)}}$ and $m_{V_{2,3}}$ being the masses of the initial and final states.
However, there still existed some issues to be clarified, e.g. the measured
large decay rates for $B_s \to \phi K^*$ and $\bar K^{*0} K^{*0}$ decays, 
the latest measurement of a smaller 
$f_L$ for $B_s \to \bar K^{*0} K^{*0}$ decay, etc.

Therefore, we would like to revisit those two-body charmless $B_s \to VV$ decays
by taking into account all currently known next-to-leading order (NLO) contributions
in the PQCD factorization approach.
We will focus on the effects of the NLO contributions arising from various possible sources,
such as the QCD vertex corrections (VC), the quark loops (QL), and the
chromomagnetic penguins~\cite{Li05,nlo05} in the SM.
As can be seen from Refs.~\cite{fan2013,xiao08b,nlo05,Li05,xiao2014},
the NLO contributions do play an important role in understanding the
known anomalies of $B$ physics such as
the amazingly large $B \to K \eta^{\prime}$ decay rates~\cite{fan2013,xiao08b},
the longitudinal-polarization dominated $B^0 \to \rho^0 \rho^0$~\cite{Li05}
and the evidently nonzero $\Delta A_{K\pi}$, i.e.,
the famous ``$K \pi$-puzzle"~\cite{nlo05,xiao2014}, and so forth.
Very recently, we extend these
calculations to the cases such as $B^0_s \to (K\pi,KK)$ decays~\cite{xiao14a},
$B^0_s \to (\pi \etap, \etap\etap)$ decays~\cite{xiao14b} and $B_s^0\to P V$ decays~\cite{xiao17}.
We found that the currently known NLO contributions can interfere with the LO part constructively
or destructively for those considered $B_s$ meson decay modes.
Consequently, the agreement between
the PQCD predictions and the experimental measurements of the
{\it CP}-averaged branching ratios, the polarization fractions and {\it CP}-violating asymmetries
was indeed improved effectively due to the inclusion of the NLO contributions.

This paper is organized as follows. In Sec.~\ref{sec:lo-nlo},
we shall
present various decay amplitudes
for the considered decay modes in the PQCD approach at the LO and NLO level.
We
show the PQCD predictions and several phenomenological analyses for the branching ratios,
{\it CP}-violating asymmetries and polarization
observables of thirteen $B_s \to VV $ decays in Sec~\ref{sec:n-d}.
A short summary is given in Sec.~\ref{sec:4}.

\section{ Decay amplitudes at LO and NLO level}\label{sec:lo-nlo}

We treat the $B_s$ meson as a heavy-light system and consider it at rest for simplicity.
By employing the light-cone coordinates, we define the $B_s$ meson with momentum
$P_1$, the emitted meson $M_2$ with the momentum $P_2$
along the direction of $n=(1,0,{\bf 0}_{\rm T})$, and the
recoiled meson $M_3$ with the momentum $P_3$ in the direction
of $v=(0,1,{\bf 0}_{\rm T})$(Here, $n$ and $v$ are the light-like
dimensionless vectors), respectively, as the following,
\beq
P_1 &=& \frac{\mbs}{\sqrt{2}} (1,1,{\bf 0}_{\rm T}), \quad
P_2 = \frac{M_{B_s}}{\sqrt{2}}(1-r_3^2,r_2^2,{\bf 0}_{\rm T}), \quad
P_3 = \frac{M_{B_s}}{\sqrt{2}} (r_3^2,1-r_2^2,{\bf 0}_{\rm T}),
\eeq
The polarization vectors of the final states can then be parametrized as:
\beq
\epsilon^L_2 &=& \frac{1}{\sqrt{2}r_2}(1-r_3^2,-r_2^2,{\bf 0}_{\rm T})   \;,
\qquad  \epsilon^L_3 =\frac{1}{\sqrt{2}r_3}(-r_3^2,1-r_2^2,{\bf 0}_{\rm T})   \;, \non
\epsilon^T_2 &=& (0,0,{\bf 1}_{\rm T})   \;,  \qquad \qquad \qquad\qquad \epsilon^T_3 =(0,0,{\bf 1}_{\rm T})   \;.
\eeq
with $\epsilon^{L(T)}$ being the longitudinal(transverse) polarization vector.

The
momenta $k_i(i=1,2,3)$ carried by the
light anti-quark in the initial $B_s$ and final
$V_{2,3}$ mesons are chosen as follows:
\beq
k_1 &=& (x_1,0,{\bf k}_{\rm 1T}), \quad
k_2 = (x_2(1-r_3^2),x_2r_2^2,{\bf k}_{\rm 2T}), \quad
k_3 = (x_3r_3^2, x_3(1-r_2^2),{\bf k}_{\rm 3T}),
\eeq
The integration over $k_{1,2}^-$ and $k_3^+$  will lead conceptually to the
decay amplitudes in the PQCD approach,
\beq
\cala(B_s^0\to V_2V_3) &\sim & \int\!\! d x_1 d x_2 d x_3 b_1 d b_1 b_2 d b_2 b_3 d b_3 \non
&& \times
\mathrm{Tr}\left [ C(t) \Phi_{B_s}(x_1,b_1) \Phi_{V_2}(x_2,b_2) \Phi_{V_3}(x_3, b_3) H(x_i,
b_i, t) S_t(x_i)\, e^{-S(t)} \right ], \quad \label{eq:a2}
\eeq
in which, $b$ is the conjugate space coordinate of transverse momentum $k_{\rm T}$,
$C(t)$ stands for the Wilson coefficients evaluated at the scale $t$,
and $\Phi$ denotes the hadron
wave functions, which are nonperturbative but universal inputs, of the
initial and final states. The kernel $H(x_i,b_i,t)$ describes the hard dynamics
associated with the effective "six-quark interaction" exchanged by a hard gluon.
The Sudakov factors $e^{-S(t)}$ and $S_t(x_i)$ together suppress the soft
dynamics in the endpoint region effectively~\cite{li2003}.

\subsection{ Wave functions and decay amplitudes}\label{sec:wf}

Without the endpoint singularities in the evaluations,
the hadron wave functions are the only input in the PQCD approach.
These nonperturbative quantities are process independent and could
be obtained with the techniques of QCD sum rule and/or Lattice QCD,
or be fitted to the measurements with good precision.

For $B_s$ meson, its wave function could be adopted with the
Lorentz structure~\cite{ali07,pqcd2}
\beq
\Phi_{B_s}&=& \frac{1}{\sqrt{6}} (\psl_{B_s} +m_{B_s}) \gamma_5 \phi_{B_s} ({\bf k}),
\label{eq:bsmeson}
\eeq
in which the distribution amplitude $\phi_{B_s}$ is modeled as
\beq
\phi_{B_s}(x,b)&=& N_{B_s} x^2(1-x)^2 \exp \left  [ -\frac{m_{B_s}^2\ x^2}{2 \omega_{B_s}^2}
-\frac{1}{2} (\omega_{B_s} b)^2\right],
\label{phib}
\eeq
with $\omega_{B_s}$ being the shape parameter.
We take $\omega_{B_s} =0.50 \pm 0.05$ GeV for the $B_s$ meson based on the
studies of lattice QCD and light-cone sum rule~\cite{wbs1,wbs2,wbs3}.
The normalization factor $N_{B_s}$ will be determined through the
normalization condition: $\int \phi_{B_s}(x,b=0)  d x =f_{B_s}/(2\sqrt{6})$ with the decay constant $f_{B_s}=0.23$ GeV.

For the vector meson, the longitudinally and transversely polarized wave functions up to twist-3 are given by~\cite{wbs3,pball98}
\begin{eqnarray}
&&\Phi_{V}^{L}\,=\,\frac{1}{\sqrt{6}}\left[m_{V}\makebox[0pt][l]{/}
\epsilon_{L}\phi_{V}(x)\,+\,\makebox[0pt][l]{/}\epsilon_{L}\makebox[-1.5pt][l]
{/}P\phi_{V}^{t}(x)+m_{V}\phi_{V}^{s}(x)\right]\nonumber\\
&&\Phi_{V}^{\perp}\,=\,\frac{1}{\sqrt{6}}\left[m_{V}\makebox[0pt][l]{/}
\epsilon_{T}\phi_{V}^{v}(x)\,+\,\makebox[0pt][l]{/}\epsilon_{T}\makebox[-1.5pt][l]
{/}P\phi_{V}^{T}(x)\,+\,m_{V}i\epsilon_{\mu\nu\rho\sigma}\gamma_{5}
\gamma^{\mu}\epsilon_{T}^{\nu}n^{\rho}v^{\sigma}\phi_{V}^{a}(x)\right],
\end{eqnarray}
where $P$ and $m_V$ are the momentum and the mass of the light vector mesons,
and  $ \epsilon_{L(T)} $ is the corresponding longitudinal(transverse)
polarization vector~\cite{Li2002}.
Here $\epsilon_{\mu\nu\rho\sigma}$ is Levi-Civita tensor with the convention $\epsilon^{0123}=1$.

The twist-2 distribution amplitudes $\phi_V(x)$ and $\phi_V^T(x)$
can be written in the following form \cite{wbs3,pball98}
\begin{eqnarray}
&&\phi_{V}(x)\,=\,\frac{3f_{V}}{\sqrt{6}}x(1-x)\left[1+a^{\|}_{1V}C_{1}^{3/2}(t)
+a_{2V}^{\|}C_2^{3/2}(t)\right],\\
&&\phi_{V}^{T}(x)\,=\,\frac{3f_{V}^T}{\sqrt{6}}x(1-x)\left[1+a^{\perp}_{1V}C_{1}^{3/2}(t)
+a_{2V}^{\perp}C_2^{3/2}(t)\right],
\end{eqnarray}
where $t=2x-1$, $f^{(T)}_{V}$  is the decay constants of the vector meson with longitudinal(transverse)  polarization.
The Gegenbauer moments here are  the same as those in Refs. \cite{wbs3,pball07,pball98}:
\begin{eqnarray}
&&a_{1\rho}^{\|(\perp)}=a_{1\omega}^{\|(\perp)}=a_{1\phi}^{\|(\perp)}=0,
\;\;a_{1K^*}^{\|(\perp)}=0.03\pm0.02\,(0.04\pm0.03)~,\nonumber\\
&&a_{2\rho}^{\|(\perp)}=a_{2\omega}^{\|(\perp)}=0.15\pm0.07\,(0.14\pm0.06)\;
a_{2\phi}^{\|(\perp)}=0\;(0.20\pm0.07)~,\nonumber\\
&&a_{2K^*}^{\|(\perp)}=0.11\pm0.09\;(0.10\pm0.08)~.
\label{eq:Geb}
\end{eqnarray}

For the twist-3 distribution amplitudes , for simplicity,
we adopt the asymptotic forms \cite{ali07,pqcd2}
\begin{eqnarray}
&&\phi_{V}^t(x)=\frac{3f_V^T}{2\sqrt{6}}t^2,\;\;\phi_{V}^s(x)=\frac{3f_{V}^T}{2\sqrt{6}}(-t),\nonumber\\
&&\phi_V^v(x)=\frac{3f_V}{8\sqrt{6}}(1+t^2),\;\;\phi_V^a(x)=\frac{3f_V}{4\sqrt{6}}(-t).
\end{eqnarray}

The above choices of vector-meson distribution amplitudes
can essentially explain the polarization fractions of the
measured $B \to K^* \phi$, $B \to K^* \rho$ and
$B \to \rho \rho$ decays\cite{lipa,lipa1,pqcdv},
together with the right branching ratios.

\subsection{ Example of the LO decay amplitudes}\label{sec:lo-aml}

In the SM,
for the considered $\bar{B}^0_s \to V V$ decays induced by
the $b \to q$ transition with $q=(d,s)$,
the weak effective Hamiltonian $H_{eff}$ can be written as\cite{buras96},
\beq
\label{eq:heff}
H_{eff} &=& \frac{G_{F}}{\sqrt{2}}     \Bigg\{ V_{ub} V_{uq}^{\ast} \Big[
 C_{1}({\mu}) O^{u}_{1}({\mu})  +  C_{2}({\mu}) O^{u}_{2}({\mu})\Big]
  -V_{tb} V_{tq}^{\ast} \Big[{\sum\limits_{i=3}^{10}} C_{i}({\mu}) O_{i}({\mu})
  \Big ] \Bigg\} + \mbox{h.c.}
\eeq
where the Fermi constant $G_{F}=1.166 39\times 10^{-5}$ GeV$^{-2}$, and
$V_{ij}$ is the Cabbibo-Kobayashi-Maskawa(CKM) matrix element,
$C_i(\mu)$ are the Wilson coefficients and $O_i(\mu)$
are the local four-quark operators. For convenience, the combinations $a_i$ of the
Wilson coefficients are defined as usual~\cite{ali07,pqcd2}:
\begin{eqnarray}
\label{eq:ai}
&&a_{1}=C_{2}+C_{1}/3,\;\;\;\;\;\;a_{2}=C_{1}+C_{2}/3,\nonumber\\
&&a_{i}=C_{i}+C_{i\pm 1}/3,\,(i=3 - 10) \;,
\end{eqnarray}
where the upper(lower) sign applies, when $i$ is odd(even).

\begin{figure}[tb]
\vspace{-2cm}
\centerline{\epsfxsize=17cm \epsffile{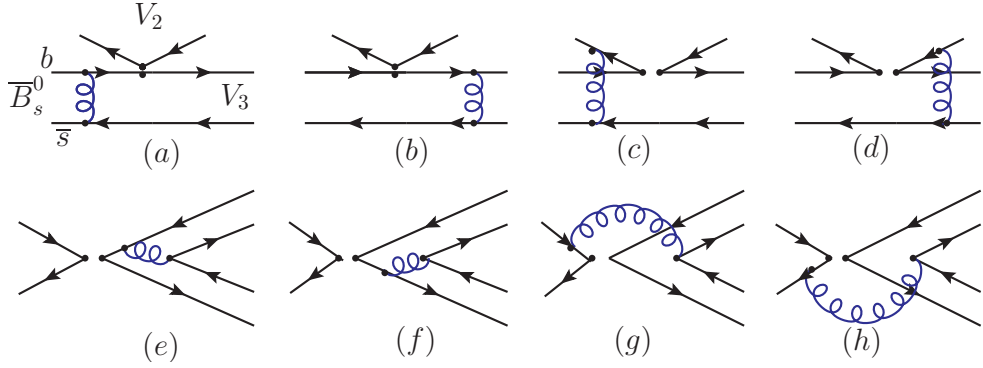}}
\vspace{-17cm}
\caption{ Typical Feynman diagrams of
$\bar{B}^0_s \to VV$ decays at leading order. }
\label{fig:fig1}
\end{figure}

At leading order, as illustrated in Fig.~\ref{fig:fig1}, there are eight types of Feynman diagrams
contributing to the $\bar{B}_s^0 \to VV$ decays,
which can be classified into three types:
the factorizable emission diagrams ( Fig.~\ref{fig:fig1}(a) and \ref{fig:fig1}(b));
the nonfactorizable emission 
diagrams (Fig.~\ref{fig:fig1}(c) and \ref{fig:fig1}(d));
and the annihilation diagrams (Fig.~\ref{fig:fig1}(e)-\ref{fig:fig1}(h)).
As mentioned in the Introduction,
the considered thirteen $\bar{B}_s^0 \to VV$ modes have been studied
at LO in the PQCD approach~\cite{ali07,pqcd2}.
The factorization formulas of decay amplitudes with various topologies
have been presented explicitly in
Ref.~\cite{ali07}. Therefore, after the confirmation by our independent
recalculations, we shall not collect those analytic expressions here for simplicity.
In this work, we aim to examine the effects of
all currently known NLO contributions to the considered $\bar{B}_s^0 \to VV$ decay modes in the PQCD
approach to see whether one can improve the consistency between
the theory and the experiment in the SM or not, which would be help for us to
judge the necessity of the exotic new physics beyond the SM.

For $\bar{B}^0_s \to VV$ decays, both of the longitudinal and
transverse polarizations will contribute.
Then, the decay amplitudes can be decomposed into three parts\cite{pqcd2}:
\begin{eqnarray}
\mathcal {A}(\epsilon_{2},\epsilon_{3})=i\mathcal
{A}^{L}+i(\epsilon_{2}^{T}\cdot\epsilon_{3}^{T})\mathcal
{A}^{N}+(\epsilon_{\mu\nu\alpha\beta}n^{\mu}v^{\nu}\epsilon_{2}^{T\alpha}
\epsilon_{3}^{T\beta})\mathcal
{A}^{T},
\label{eq:amplitude}
\end{eqnarray}
where $\mathcal{A}^{L},\mathcal{A}^{N}$ and $\mathcal{A}^{T}$ correspond to the longitudinally,
normally and transversely polarized amplitudes, respectively,
whose detailed expressions can be inferred from Refs.~\cite{ali07,pqcd2}.

\subsection{ NLO contributions}
\begin{figure}[tb]
\vspace{-2cm} \centerline{\epsfxsize=17 cm \epsffile{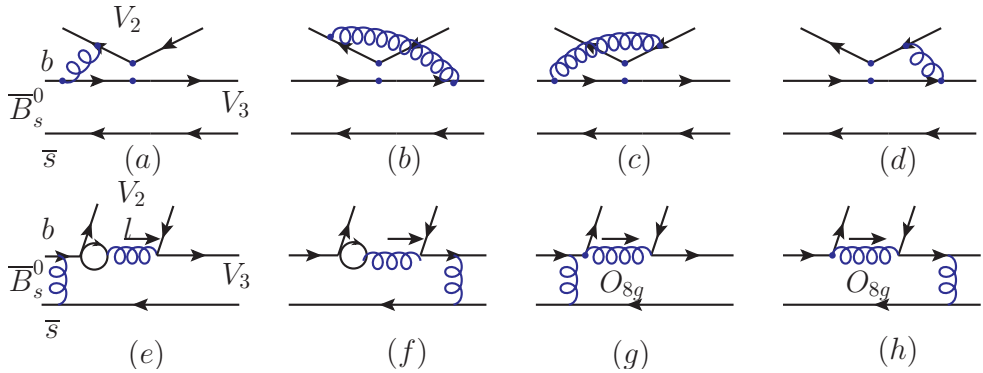}}
\vspace{-17cm}
\caption{Feynman diagrams for NLO contributions:  the vertex corrections (a-d);
the quark-loop contributions (e-f) and the chromomagnetic penguin contributions (g-h). }
\label{fig:fig2}
\end{figure}

In the framework of the PQCD approach, many two-body charmless $B/B_s \to PP, PV$ decays have been
investigated by including currently known NLO contributions,for example,
in Refs.~\cite{fan2013,nlo05,xiao14a,xiao14b,xiao17,xiao08b,xiao2014,zhang09,zhou12}.
Of course, some NLO contributions are still not known at present, as discussed in Ref.~\cite{fan2013}.
The currently known NLO corrections to the LO PQCD predictions
of $B_s \to VV$ decays are the following:
\begin{itemize}
\item[]{(a)}
The NLO Wilson coefficients $C_i(m_W)$(NLO-WC), the renormalization group running matrix $U(m_1,m_2,\alpha)$ at NLO level
and the strong coupling constant $\alpha_s(\mu)$ at two-loop level as presented in Ref.~\cite{buras96};

\item[]{(b)}
The NLO contributions from the vertex corrections (VC)~\cite{Li05,nlo05}
as illustrated in Figs.~\ref{fig:fig2}(a)-\ref{fig:fig2}(d);

\item[]{(c)}
The NLO contributions from the quark-loops (QL) ~\cite{nlo05,Li05}
as shown in Figs.~\ref{fig:fig2}(e)-\ref{fig:fig2}(f);

\item[]{(d)}
The NLO contributions from the chromo-magnetic penguin (MP) operator $O_{8g}$ \cite{o8g2003,Li05,nlo05} as illustrated
in Figs.~\ref{fig:fig2}(g)-\ref{fig:fig2}(h).
\end{itemize}

In this paper, we adopt directly the formulas for all currently known
NLO contributions from Refs.~\cite{nlo05,Li05,o8g2003,xiao08b,fan2013,xiao14a,xiao14b,xiao2014}
without further discussions about the details.
Moreover, some essential comments
should be given for those still unknown NLO corrections to the nonfactorizable emission amplitudes
and the annihilation amplitudes as follows:
\begin{itemize}
\item[]{(a)}
For the nonfactorizable emission diagrams as shown in Fig.~\ref{fig:fig1},
since the hard gluons
are emitted from the upper quark line of Fig.~\ref{fig:fig1}(c) and the upper anti-quark
line of Fig.~\ref{fig:fig1}(d) respectively, the contribution from these two figures will be
strongly cancelled each other, the remaining contribution is therefore becoming rather small
in magnitude. In NLO level, another suppression factor $\alpha_s(t)$ will appear, the
resultant NLO contribution from the hard-spectators should become much smaller than
the dominant contribution from the ¡±tree¡± emission diagrams (Fig.~\ref{fig:fig1}(a) and~\ref{fig:fig1}(b)).

\item[]{(b)}
For the annihilation diagrams as shown in Fig.~\ref{fig:fig1}(e)-\ref{fig:fig1}(h),
the corresponding NLO contributions are in fact doubly suppressed by the factors $1/m_{B_s}$ and $\alpha_s(t)$,
and consequently must become much smaller than those dominant LO contribution from Fig.~\ref{fig:fig1}(a)
and~\ref{fig:fig1}(b).
\end{itemize}
Therefore, it is reasonable for us to expect that those still unknown NLO contributions in the PQCD approach
are in fact the higher order corrections to the already small LO pieces, and
should be much smaller than the dominant contribution for the considered
decays, say less than 5\% of the dominant ones.

According to Refs.~\cite{Li05,nlo05}, the vertex corrections can be absorbed
into the redefinition of the Wilson coefficients $a_i(\mu)$ by adding a vertex-function $V_i(M)$ to them.
\begin{eqnarray}
a_{1,2}(\mu) &\to& a_{1,2}(\mu) +\frac{\alpha_s(\mu)}{9\pi}\; C_{1,2}(\mu)\; V_{1,2}(M) \;, \non
a_i(\mu) &\to& a_i(\mu) +\frac{\alpha_s(\mu)}{9\pi}\; C_{i+ 1}(\mu) \; V_i(M), \qquad {\rm for\ \ } i=3,5,7,9, \non
a_j(\mu) &\to& a_j(\mu) +\frac{\alpha_s(\mu)}{9\pi}\; C_{j-1}(\mu) \; V_j(M), \qquad {\rm for \ \ } j=4,6,8,10,
\label{wnlo}
\end{eqnarray}
where $M$ denotes the vector meson emitted from the weak vertex ( i.e. the $M_2$ in Fig.~\ref{fig:fig2}(a)-\ref{fig:fig2}(d)).
The expressions of the vertex-functions $V_i(M)$ with both longitudinal and  transverse components can be found easily in Refs.~\cite{qcdfpppv,nlovc}.

The NLO ``Quark-Loop" and ``Magnetic-Penguin" contributions  are in fact
a kind of penguin corrections
with insertion of the four-quark operators and the chromo-magnetic operator
$O_{8g}$, respectively,
as shown in Figs.~\ref{fig:fig2}(e,f) and \ref{fig:fig2}(g,h).
For the $b\to s$ transition, for example, the corresponding effective Hamiltonian $H_{eff}^{ql}$ and
$H_{eff}^{mp}$ can be written in the following form:
\beq
H_{eff}^{(ql)}&=&-\sum\limits_{q=u,c,t}\sum\limits_{q{\prime}}\frac{G_F}{\sqrt{2}}
V_{qb}^{*}V_{qs}\frac{\alpha_s(\mu)}{2\pi}C^{q}(\mu,l^2)\left[ \bar{b}\gamma_\rho
\left(1-\gamma_5\right)T^as\right ]\left(\bar{q}^{\prime}\gamma^\rho
T^a q^{\prime}\right),\label{eq:heff-ql}\\
H_{eff}^{mp} &=&-\frac{G_F}{\sqrt{2}} \frac{g_s}{8\pi^2}m_b\;
V_{tb}^*V_{ts}\; C_{8g}^{eff} \; \bar{s}_i \;\sigma^{\mu\nu}\; (1+\gamma_5)\;
 T^a_{ij}\; G^a_{\mu\nu}\;  b_j, \label{eq:heff-o8g}
\eeq
where $l^2$ is  the invariant mass of the gluon which attaches the quark loops
in Figs.~\ref{fig:fig2}(e,f), and the functions $C^{q}(\mu,l^2)$ can be inferred from
Refs.~\cite{Li05,nlo05,fan2013,xiao08b,xiao2014}.
The $C_{8g}^{eff}$ in Eq.~(\ref{eq:heff-o8g}) is an effective Wilson coefficient
with the definition of $C_{8g}^{eff}= C_{8g} + C_5$~\cite{buras96}.

With explicit evaluations, we find the following three points:
\begin{enumerate}
\item[(1)]
For the pure annihilation decays of $B_s^0 \to \rho\rho$, $\rho\omega$ and $ \omega\omega$, they
do not receive the NLO contributions from the vertex corrections, the quark-loop
and the magnetic-penguin diagrams. The NLO correction to these decay modes
comes only from the NLO-WCs and the the strong coupling constant $\alpha_s(\mu)$ at the two-loop level.

\item[(2)]
For the $B_s^0 \to \rho^0\phi$ and $\omega \phi$ channels with only $B_s \to \phi$ transition and
no annihilation diagrams, the "quark-loop" and "magnetic-penguin" diagrams cannot contribute to these two
decay modes.
The related NLO contributions are mainly induced by the vertex corrections
to the emitted $\rho$ or $\omega$ mesons.

\item[(3)]
For the remaining seven decay modes, besides the LO decay amplitudes, all of the currently known NLO
contributions should be taken into account as follows:
\begin{eqnarray}
{\renewcommand\arraystretch{2.0} \begin{array}{ll}
\displaystyle {\cal A}^{(u),i}_{\rho^0K^{*0} } \, \to\, {\cal
A}^{(u),i}_{\rho^0K^{*0} }+{\cal M}^{(u,c),i}_{\rho^0K^{*0}}\;,
&\displaystyle {\cal A}^{(t),i}_{\rho^0K^{*0} } \, \to\, {\cal
A}^{(t),i}_{\rho^0K^{*0} }-{\cal M}^{(t),i}_{\rho^0K^{*0}}-{\cal
M}^{(g),i}_{\rho^0K^{*0}}\;,
\\
\displaystyle {\cal A}^{(u),i}_{\rho^-K^{*+} } \, \to\, {\cal
A}^{(u),i}_{\rho^-K^{*+} }+{\cal M}^{(u,c),i}_{\rho^-K^{*+}}\;,
&\displaystyle {\cal A}^{(t),i}_{\rho^-K^{*+} } \, \to\, {\cal
A}^{(t),i}_{\rho^-K^{*+} }-{\cal M}^{(t),i}_{\rho^-K^{*+}}-{\cal
M}^{(g),i}_{\rho^-K^{*+}}\;,
\\
\displaystyle {\cal A}^{(u),i}_{\omega K^{*0} } \, \to\, {\cal
A}^{(u),i}_{\omega K^{*0} }+{\cal M}^{(u,c),i}_{\omega K^{*0}}\;,
&\displaystyle {\cal A}^{(t),i}_{\omega K^{*0} } \, \to\, {\cal
A}^{(t),i}_{\omega K^{*0} }-{\cal M}^{(t),i}_{\omega K^{*0}}-{\cal
M}^{(g),i}_{\omega K^{*0}}\;,
\\
\displaystyle {\cal A}^{(u),i}_{\phi K^{*0} } \, \to\, {\cal
A}^{(u),i}_{\phi K^{*0} }+{\cal M}^{(u,c),i}_{\phi K^{*0}}\;,
&\displaystyle {\cal A}^{(t),i}_{\phi K^{*0} } \, \to\, {\cal
A}^{(t),i}_{\phi K^{*0} }-{\cal M}^{(t),i}_{\phi K^{*0}}-{\cal
M}^{(g),i}_{\phi K^{*0}}\;,
\\
\displaystyle {\cal A}^{(u),i}_{K^{*-}K^{*+} } \, \to\, {\cal
A}^{(u),i}_{K^{*-}K^{*+} }+{\cal M}^{(u,c),i}_{K^{*-}K^{*+}}\;,
&\displaystyle {\cal A}^{(t),i}_{K^{*-}K^{*+} } \, \to\, {\cal
A}^{(t),i}_{K^{*-}K^{*+} }-{\cal M}^{(t),i}_{K^{*-}K^{*+}}-{\cal
M}^{(g),i}_{K^{*-}K^{*+}}\;,
\\
\displaystyle {\cal A}^{(u),i}_{K^{*0}\bar K^{*0} } \, \to\, {\cal
A}^{(u),i}_{K^{*0}\bar K^{*0} }+{\cal M}^{(u,c),i}_{K^{*0}\bar K^{*0}}\;,
&\displaystyle {\cal A}^{(t),i}_{K^{*0}\bar K^{*0} } \, \to\, {\cal
A}^{(t),i}_{K^{*0}\bar K^{*0} }-{\cal M}^{(t),i}_{K^{*0}\bar K^{*0}}-{\cal
M}^{(g),i}_{K^{*0}\bar K^{*0}}\;,
\\
\displaystyle {\cal A}^{(u),i}_{\phi \phi } \, \to\, {\cal
A}^{(u),i}_{\phi \phi }+{\cal M}^{(u,c),i}_{\phi \phi}\;,
&\displaystyle {\cal A}^{(t),i}_{\phi \phi } \, \to\, {\cal
A}^{(t),i}_{\phi \phi }-{\cal M}^{(t),i}_{\phi \phi}-{\cal
M}^{(g),i}_{\phi \phi}\;,
\end{array} }
\end{eqnarray}
where $i=L,N,T$ and the terms ${\cal A}_{V_2 V_3}^{(u,t),i}$ stand for
the LO amplitudes, while ${\cal M}_{V_2 V_3}^{(u, c, t ),i}$
and ${\cal M}_{V_2 V_3}^{(g),i}$ are the NLO ones, which describe the NLO
contributions arising from the up-loop, charm-loop,
QCD-penguin-loop, and magnetic-penguin diagrams, respectively.

\end{enumerate}

Now, we can calculate the decay amplitudes ${\cal M}_{V_2 V_3}^{(ql),i}$ and ${\cal M}_{V_2 V_3}^{(mp),i}$ in the
PQCD approach.
As mentioned in Eq.~(\ref{eq:amplitude}), for $B_s \to VV $ decays, there are
three individual polarization amplitudes ${\cal M}_{V_2 V_3}^{L,N,T}$.
For the longitudinal components, the NLO decay amplitudes ${\cal M}_{V_2 V_3}^{(ql),L}$ and ${\cal M}_{V_2 V_3}^{(mp),L}$
can be written as:
\beq
{\cal M}^{(ql),L}_{V_2 V_3}&=& -8m_{B_s}^4\frac{C_F^2}{\sqrt{6}} \int_0^1 dx_1dx_2dx_3
\int_0^\infty b_1db_1b_3db_3 \,\phi_{B_s}(x_1, b_1) \Bigl \{ \left [(1+x_3)\phi_2(x_2) \phi_3(x_3)
\right.
\non && \left. -2 r_2\phi_2^s(x_2) \phi_3(x_3)+ r_3(1-2x_3)\phi_2(x_2)(\phi_3^s(x_3)
+ \phi_3^t(x_3)) -2r_2r_3\phi_2^s(x_2)((2+x_3)\phi_3^s(x_3)\right.
\non &&
-x_3\phi_3^t(x_3))]\cdot \alpha_s^2(t_a) \cdot h_e(x_1,x_3,b_1,b_3)\cdot  \exp\left [-S_{ab}(t_a)\right ]\; C^{(q)}(t_a,l^2)+ [2r_3\phi_2(x_2)\phi_3^s(x_3)
\non &&
-4r_2r_3\phi_2^s(x_2)]\phi_3^s(x_3) \cdot
\alpha_s^2(t_b) \cdot h_e(x_3,x_1,b_3,b_1)
\cdot \exp[-S_{ab}(t_b)] \; C^{(q)}(t_b,l'^2)\Bigr \},
\eeq
\beq
{\cal M}^{(mp),L}_{V_2V_3} &=& 8m_{B_s}^6\frac{C_F^2}{\sqrt{6}} \int_0^1 dx_1dx_2dx_3
\int_0^\infty b_1db_1b_2db_2b_3db_3\, \phi_{B_s}(x_1, b_1)\non
&&\times \left \{ \left [ (-1+x_3) \left [ 2\phi_3(x_3)- r_3(x_3-1)\phi_3^t(x_3)
+r_3 (x_3+3)\phi_3^s(x_3)\right ] \phi_2(x_2) \right.\right. \non
&&  \left.\left.
+r_2 x_2(3\phi_2^s(x_2)-\phi_2^t(x_2))\left [(1+x_3) \phi_3(x_3)-r_3(2x_3-1)(\phi_3^s(x_3)+\phi_3^t(x_3)) \right ]
\right.\right.
\non &&
\left. +r_2r_3 (x_3-1)(3\phi_2^s(x_2)+\phi_2^t(x_2)) (\phi_3^t(x_3) -\phi_3^s(x_3) ) \right ] \non
&&  \hspace{0.5cm} \cdot \alpha_s^2(t_a) h_g(x_i,b_i)\cdot \exp[-S_{cd}(t_a)]\; C_{8g}^{eff}(t_a)
\non
&&
-\left [ 4r_3\phi_2(x_2)-2r_2r_3 x_2(3\phi_2^s(x_2)-\phi_2^t(x_2)) \right ] \phi_3^s(x_3) \non
&&
\left. \hspace{0.5cm}\cdot \alpha_s^2(t_b) \cdot h'_g(x_i,b_i)\cdot \exp[-S_{cd}(t_b) ] \cdot C_{8g}^{eff}(t_b )\right \}.
 \eeq
with $C_F=4/3$.

The transverse components ${\cal M}_{V_2 V_3}^{(ql),N,T}$ and
${\cal M}_{V_2 V_3}^{(mp),N,T}$ of the corresponding decay amplitudes
can be written in the form of
\beq
{\cal M}^{(ql),N}_{V_2 V_3}&=& -8m_{B_s}^4 r_2 \frac{C_F^2}{\sqrt{6}} \int_0^1 dx_1dx_2dx_3
\int_0^\infty b_1db_1b_3db_3 \,\phi_{B_s}(x_1, b_1) \Bigl \{ \left [\phi^T_3(x_3)(\phi_2^a(x_2)+\phi_2^v(x_2))
\right.
\non && \left. + r_3(x_3+2)(\phi_2^a(x_2)\phi_3^a(x_3)+\phi_2^v(x_2)\phi_3^v(x_3))
- r_3x_3(\phi_2^a(x_2)\phi_3^v(x_3)+\phi_2^v(x_2)\phi_3^a(x_3))\right.]
\non &&
 \hspace{0.5cm}\cdot \alpha_s^2(t_a) \cdot h_e(x_1,x_3,b_1,b_3)\cdot  \exp\left [-S_{ab}(t_a)\right ]\; C^{(q)}(t_a,l^2)\non
&&
+ \left [ r_2r_3(\phi^a_2(x_2)\phi_3^a(x_3)+\phi^a_2(x_2)\phi_3^v(x_3)
+\phi^v_2(x_2)\phi_3^a(x_3)+\phi^v_2(x_2)\phi_3^v(x_3)) \right ] \phi_3^s(x_3) \non
&& \hspace{0.5cm}\cdot \alpha_s^2(t_b) \cdot h_e(x_3,x_1,b_3,b_1)
\cdot \exp[-S_{ab}(t_b)] \; C^{(q)}(t_b,l'^2)\Bigr \},
\eeq
\beq
{\cal M}^{(ql),T}_{V_2 V_3}&=& -8m_{B_s}^4 r_2 \frac{C_F^2}{\sqrt{6}} \int_0^1 dx_1dx_2dx_3
\int_0^\infty b_1db_1b_3db_3 \,\phi_{B_s}(x_1, b_1) \Bigl \{ \left [\phi^T_3(x_3)(\phi_2^a(x_2)+\phi_2^v(x_2))
\right.
\non && \left. + r_3(x_3+2)(\phi_2^a(x_2)\phi_3^v(x_3)+\phi_2^v(x_2)\phi_3^a(x_3))
- r_3x_3(\phi_2^a(x_2)\phi_3^a(x_3)+\phi_2^v(x_2)\phi_3^v(x_3))\right.]
\non &&
\hspace{0.5cm}\cdot \alpha_s^2(t_a) \cdot h_e(x_1,x_3,b_1,b_3)\cdot  \exp\left [-S_{ab}(t_a)\right ]\; C^{(q)}(t_a,l^2)\non
&&
+ \left [ r_2r_3(\phi^a_2(x_2)\phi_3^a(x_3)+\phi^a_2(x_2)\phi_3^v(x_3)
+\phi^v_2(x_2)\phi_3^a(x_3)+\phi^v_2(x_2)\phi_3^v(x_3)) \right ]\phi_3^s(x_3)\non
&&
\hspace{0.5cm}\cdot \alpha_s^2(t_b) \cdot h_e(x_3,x_1,b_3,b_1)
\cdot \exp[-S_{ab}(t_b)] \; C^{(q)}(t_b,l'^2)\Bigr \},
\eeq
\beq
{\cal M}^{(mp),N}_{V_2V_3} &=& 8m_{B_s}^6\frac{C_F^2}{\sqrt{6}} \int_0^1 dx_1dx_2dx_3
\int_0^\infty b_1db_1b_2db_2b_3db_3\, \phi_{B_s}(x_1, b_1)\non
&&\times \left \{ \left [-r_3(x_3^2-1)\phi_2^T(x_2)(\phi_3^a(x_3)-\phi_3^v(x_3))
-r_2x_2(1+x_3)\phi_3^T(x_3)(\phi_2^a(x_2)-\phi_2^v(x_2)) \right.\right. \non
&&
+r_2 r_3(2 x_2 x_3-x_2+x_3-1)(\phi_2^a(x_2)\phi_3^a(x_3)+\phi_2^v(x_2)\phi_3^v(x_3))\non
&& \left. +r_2 r_3(2 x_2 x_3-x_2 -x_3-1) (\phi_2^a(x_2)\phi_3^v(x_3)+\phi_2^v(x_2)\phi_3^a(x_3)) \right ]\non
&& \hspace{0.5cm} \cdot \alpha_s^2(t_a) h_g(x_i,b_i)\cdot \exp[-S_{cd}(t_a)]\; C_{8g}^{eff}(t_a) \non
&&
-\left [r_2 r_3 x_2(\phi_2^a(x_2)+\phi_2^v(x_2))(\phi_3^a(x_3)+\phi_3^v(x_3)) \right ]  \non
&& \left. \hspace{0.5cm} \cdot \alpha_s^2(t_b) \cdot h'_g(x_i,b_i)\cdot \exp[-S_{cd}(t_b) ] \cdot C_{8g}^{eff}(t_b)\right \},
\eeq
\beq
{\cal M}^{(mp),T}_{V_2V_3} &=& 8m_{B_s}^6\frac{C_F^2}{\sqrt{6}} \int_0^1 dx_1dx_2dx_3
\int_0^\infty b_1db_1b_2db_2b_3db_3\, \phi_{B_s}(x_1, b_1)\non
&&\times \left \{ \left [-r_3(x_3^2-1)\phi_2^T(x_2)(\phi_3^a(x_3)-\phi_3^v(x_3))
-r_2x_2(1+x_3)\phi_3^T(x_3)\left (\phi_2^a(x_2)-\phi_2^v(x_2) \right ) \right.\right. \non
&&
+r_2 r_3(2 x_2 x_3-x_2+x_3-1)\left (\phi_2^a(x_2)\phi_3^v(x_3)+\phi_2^v(x_2)\phi_3^a(x_3) \right ) \non
&& \left. +r_2 r_3\left ( 2 x_2 x_3-x_2-x_3-1 \right )\left ( \phi_2^a(x_2)\phi_3^a(x_3)+\phi_2^v(x_2)\phi_3^v(x_3) \right )
\right ] \non
&&  \hspace{0.5cm}\cdot \alpha_s^2(t_a) h_g(x_i,b_i)\cdot \exp[-S_{cd}(t_a)]\; C_{8g}^{eff}(t_a)
\non
&&
-\left [ r_2 r_3 x_2(\phi_2^a(x_2)+\phi_2^v(x_2))\left ( \phi_3^a(x_3)+\phi_3^v(x_3) \right ) \right ]\non
&& \left. \hspace{0.5cm}\cdot \alpha_s^2(t_b) \cdot h'_g(x_i,b_i)\cdot \exp[-S_{cd}(t_b) ] \cdot C_{8g}^{eff}(t_b)\right \},
 \eeq
In the above equations,
the explicit expressions for the hard functions $(h_e,h_g,h'_g)$, the functions
$C^{(q)}(t_a,l^2)$ and $C^{(q)}(t_b,l'^2)$, the Sudakov factors $S_{ab}(t)$ and $S_{cd}(t)$,
the hard scales $t_{a,b}$ and the effective Wilson coefficients $C_{8g}^{eff}(t)$,
can be found easily, for example, in Refs.~\cite{Li05,nlo05,fan2013,xiao08b,xiao14a,xiao14b,xiao2014}.

\section{Numerical results}\label{sec:n-d}

In the numerical calculations, the following input parameters will be used implicitly.
The masses, decay constants and QCD scales are in units of GeV \cite{pqcd2,pdg2016}
\beq
\Lambda_{\overline{\mathrm{MS}}}^{(f=5)} &=& 0.225,
\quad  f_{B_{s}} = 0.23, \quad f_{\rho}=0.216,\quad f_{\rho}^{T}=0.165,
\quad f_{\omega}=0.187, \quad f_{\omega}^{T}=0.151, \non
M_{B_{s}} &=&  5.37,\quad f_{\phi}=0.215, \quad f_{\phi}^{T}=0.186,
\quad f_{K^*}=0.220, \quad f_{K^*}^{T}=0.185, \quad m_{K^*}=0.892,\non
m_{\phi}&=&1.02,\quad m_{\rho}=0.77, \quad m_{\omega}= 0.78, \quad \tau_{B^0_s}=1.497 {\rm ps},
\quad m_b=4.8, \quad M_W=80.42.
 \label{eq:para}
\eeq

For the CKM matrix elements, we adopt the Wolfenstein parametrization up to
${\cal O}(\lambda^5)$ with the updated parameters as~\cite{pdg2016}
\beq
\lambda=0.22537\pm 0.00061, \quad A=0.814^{+0.023}_{-0.024}, \quad\bar \rho=0.117 \pm 0.021, \quad\bar \eta=0.353\pm 0.013.
\label{eq:ckm}
\eeq

The total decay amplitude for $\bar{B}_s^0 \to VV$ decays can be expressed as
\begin{eqnarray}
\mid\mathcal{M}(\bar{B}_s^0\rightarrow
f)\mid^2\,=\, \mid\mathcal{A}_{0}\mid^{2}+\mid\mathcal{A}_{\|}\mid^{2}+\mid\mathcal{A}_{\bot}\mid^{2}
\end{eqnarray}
where  $\mathcal{A}_{0},\mathcal{A}_{\|},\mathcal{A}_{\bot}$
denote the longitudinal, parallel, and perpendicular polarization amplitude in the transversity basis, respectively,
which are defined as follows~\cite{pqcd2}:
\begin{eqnarray}
A_0=\mathcal{A}_{L},\quad A_{\parallel}=\sqrt{2}\mathcal{A}^{N},\quad
A_{\perp}=\sqrt{2}\mathcal{A}^{T}.
\end{eqnarray}
Therefore, the CP-averaged branching ratio can be written as
\begin{eqnarray}
{ BR}&=&\frac{|\bf{P}|}{16\pi
M_{B}^{2}} \tau_{B_s}\left [ \mid\mathcal{M}(\bar{B}_s^0\rightarrow f)\mid^2+
\mid\mathcal{M}(B_s^0\rightarrow \bar{f})\mid^2\right ],
\end{eqnarray}
where $\bf{P}$ is the 3-momentum of either of the two vector mesons in the final state and $\tau_{B_s}$
is the lifetime of the $B_s$ meson.

The definitions of the polarization fractions $f_{L,\|,\bot}$ and the relative phases $\phi_{\|, \bot}$
are given as
\begin{eqnarray}
f_{L,\parallel,\perp}=\frac{\mid A_{L,\parallel,\perp}\mid^2}{\mid A_{0}\mid^{2}+\mid A_{\parallel}\mid^{2}+\mid A_{\perp}\mid^{2}},\;\;\;
\;\;\;\phi_{\parallel,\bot}={\rm Arg}(A_{\parallel,\bot}/A_{0}).
\end{eqnarray}

Combined with the CP-conjugated decay, the direct CP asymmetry can be defined as
\begin{eqnarray}
A^{\rm dir}_{CP}&=&\frac{BR(\bar{B}_s^0\rightarrow f)-BR(B_s^0\rightarrow \bar{f})}
{BR(\bar{B}_s^0\rightarrow f)+BR(B_s^0\rightarrow \bar{f})} \non
&=&\frac{\mid\mathcal{M}(\bar{B}_s^0\rightarrow f)\mid^2-\mid\mathcal{M}(B_s^0\rightarrow \bar{f})\mid^2}
{\mid\mathcal{M}(\bar{B}_s^0\rightarrow f)\mid^2+\mid\mathcal{M}(B_s^0\rightarrow \bar{f})\mid^2}.
\end{eqnarray}
Besides, we also evaluate the following observables:
\begin{eqnarray}
A_{CP}^0&=&\frac{f_L-\bar{f}_L}{f_L+\bar{f}_L},\;\;\;\;A_{CP}^{\bot}=\frac{f_{\bot}-\bar{f}_{\bot}}{f_{\bot}+\bar{f}_{\bot}},\non
\Delta\phi_{\parallel}&=&\frac{\phi_{\parallel}-\bar{\phi}_{\parallel}}{2},\;\;\;\;
\Delta\phi_{\bot}=\frac{\phi_{\bot}-\bar{\phi}_{\bot}}{2}.
\end{eqnarray}

\begin{table}[t]  
\caption{ The predicted branching ratios(in units of $10^{-6}$) of the
$\bar B_s^0\to \phi \phi, K^{*0} \phi$, and $\bar K^{*0} K^{*0}$ decays in the
PQCD approach at LO and NLO level.
As a comparison, the numerical results from the previous PQCD,
QCDF,SCET,and FAT approaches are also quoted.}
\label{Tab:brexp}
\begin{tabular*}{16cm}{@{\extracolsep{\fill}}l|llllll} \hline\hline
{\rm Modes}&{\rm LO} &{\rm NLOWC}&{\rm +VC}&{\rm +QL}&{\rm +MP}&{\rm NLO} \\ \hline
$\bar B_s^0\to \phi \phi$&              16.4&    $19.8 $&$17.6   $&$21.2  $&$15.2   $&$18.8^{+4.9}_{-3.8}$     \\
$\bar B_s^0\to K^{*0} \phi$&            0.38&    $0.46 $&$0.42   $&$0.60  $&$0.34   $&$0.42^{+0.13}_{-0.10}$   \\
$\bar B_s^0\to \bar K^{*0} K^{*0}$&     5.0&     $6.61 $&$7.16  $&$8.68  $&$4.70   $&$6.68^{+2.9}_{-2.2}$\\
 \hline
{\rm Modes}&{\rm PQCD\cite{pqcd2}} &{\rm QCDF\cite{qcdf09}}&{\rm QCDF\cite{qcdf07}}&{\rm SCET\cite{scet}}&{\rm FAT\cite{fat}}
&{\rm Data\cite{lhcbphi2,lhcbphiks,lhcbks1}} \\ \hline
$\bar B_s^0\to \phi \phi$&              $16.7^{+4.9}_{-3.8}$&     $16.7^{+11.9}_{-9.1} $&$21.8^{+30.4}_{-17.1}   $&$19.0\pm 6.5  $&$26.4\pm7.6  $&$18.4\pm1.9$     \\
$\bar B_s^0\to K^{*0} \phi$&            $0.39^{+0.20}_{-0.17}$&   $0.37^{+0.25}_{-0.21} $&$0.4^{+0.51}_{-0.31}   $&$0.56\pm0.19  $&$0.7\pm0.18  $&$1.13\pm 0.30$   \\
$\bar B_s^0\to \bar K^{*0} K^{*0}$&     $5.4^{+3.0}_{-2.4}$&      $6.6\pm 2.2         $&$9.1^{+11.3}_{-6.8} $&$8.6\pm3.1 $&$14.9\pm3.6      $&$10.8\pm 2.84$\\
\hline\hline
\end{tabular*}
\end{table}
\begin{table}[]  
\caption{ Same as Table \ref{Tab:brexp} but for the longitudinal $f_L$ (the first entry)
and perpendicular $f_{\bot}$(the second entry) polarization fractions($\%$).}
\label{Tab:fexp}
\begin{tabular*}{16cm}{@{\extracolsep{\fill}}l|llllll} \hline\hline
{\rm Modes}&{\rm LO} &{\rm NLOWC}&{\rm +VC}&{\rm +QL}&{\rm +MP}&{\rm NLO} \\ \hline
$\bar B_s^0\to \phi \phi$&              $31.2$&    $37.3 $&$27.0   $&$50.4  $&$23.9   $&$31.6^{+6.7}_{-5.3}$     \\
                                       &$33.2$&    $30.1 $&$37.9   $&$23.8  $&$36.6   $&$35.5^{+2.8}_{-4.2}$   \\

$\bar B_s^0\to K^{*0} \phi$&            $46.2$&    $53.2 $&$46.8   $&$62.8  $&$37.6   $&$47.1^{+8.2}_{-7.4}$   \\
                                       &$26.7$&    $23.1 $&$28.6   $&$18.4  $&$30.8   $&$28.3^{+3.4}_{-2.8}$   \\

$\bar B_s^0\to \bar K^{*0} K^{*0}$&     $35.7$&    $45.4 $&$49.0   $&$57.5  $&$25.2   $&$43.4^{+12.7}_{-12.9}$\\
                                       &$28.2$&    $25.4 $&$22.8   $&$19.7  $&$35.3   $&$23.5^{+5.8}_{-5.9}$   \\
\hline
{\rm Modes}&{\rm PQCD\cite{pqcd2}} &{\rm QCDF\cite{qcdf09}}&{\rm QCDF\cite{qcdf07}}&{\rm SCET\cite{scet}}
&{\rm FAT\cite{fat}}&{\rm Data\cite{cdf,lhcbphiks,lhcbks,lhcbks1}} \\
\hline
$\bar B_s^0\to \phi \phi$&              $34.7^{+8.9}_{-7.1}$&    $36.2^{+23.2}_{-18.4}  $&$43^{+1}_{-34}   $&$51\pm16.4   $&$39.7\pm16.0  $&$34.8\pm4.6$     \\
                                       &$31.6^{+3.5}_{-4.4}$&    $-    $&$-               $&$22.2\pm9.9      $&$31.2\pm8.9  $&$36.5\pm5.2$     \\
$\bar B_s^0\to K^{*0} \phi$&            $50.0^{+8.1}_{-7.2}$&    $43^{+21.1}_{-18.1}    $&$40^{+67}_{-35}  $&$54.6\pm16.0 $&$38.9\pm14.7  $&$51\pm 16.5$   \\
                                       &$24.2^{+3.6}_{-3.9}$&    $-     $&$-              $&$20.5\pm9.1      $&$31.4\pm8.1  $&$28\pm11.2$     \\
$\bar B_s^0\to \bar K^{*0} K^{*0}$&     $38.3^{+12.1}_{-10.5}$&  $56^{+22.4}_{-26.1}    $&$63^{+42}_{-29}  $&$44.9\pm18.3 $&$34.3\pm12.6  $&$20.1\pm 6.9$\\
                                       &$30.0^{+5.3}_{-6.1}$&    $-     $&$-              $&$24.9\pm11.1     $&$33.2\pm6.9  $&$38\pm11.4$     \\
\hline\hline
\end{tabular*}
\end{table}

\begin{table}[] 
\centering
\caption{The PQCD predictions for the branching ratios, $f_{L}$($\%$) and $f_{\perp}$($\%$) for
the relevant decays. The data are taken from Refs.~\cite{cdf,lhcbphiks,lhcbks,lhcbphi2,lhcbks1}.
 The meaning of the labels are described in the text. }
\vspace{0.cm}
\label{Tab:ann}
\begin{tabular*}{17cm}{@{\extracolsep{\fill}}l|lll|lll|lll}
\hline \hline
 &
 \multicolumn{3}{c|}{\rm Br $(10^{-6})$} &  \multicolumn{3}{c|}{$f_L$ $(\%)$}&  \multicolumn{3}{c}{ $f_\bot$ $(\%)$}\\
 \hline
 {\rm  Mode}&{\rm NLO}&{\rm No Ann}&{\rm Data}&{\rm NLO}&{\rm No Ann}&{\rm Data}&{\rm NLO}
 &{\rm No Ann}&{\rm Data}\\
 \hline
 $\bar B_s^0\to \phi \phi$              &$18.8^{+4.9}_{-3.8}$  &$7.6^{+4.9}_{-3.8}$    &$18.4\pm1.9$   &$31.6^{+6.7}_{-5.3}$ &$78.2^{+3.8}_{-3.1}$   &$34.8\pm4.6$
                                        &$35.5^{+2.8}_{-4.2}$  &$12.7^{+2.6}_{-3.0}$   &$36.5\pm5.2$ \\

 $\bar B_s^0\to K^{*0} \phi$            &$0.42^{+0.13}_{-0.10}$  &$0.15^{+0.05}_{-0.03}$ &$1.13\pm0.30$  &$47.1^{+8.2}_{-7.4}$ &$80.8^{+5.3}_{-5.1}$   &$51\pm16.5$
                                        &$28.3^{+3.4}_{-2.8}$    &$11.6^{+3.1}_{-3.6}$   &$28\pm11.2$\\

 $\bar B_s^0\to \bar K^{*0} K^{*0}$     &$6.68^{+2.9}_{-2.2}$   &$4.86^{+1.5}_{-1.2}$   &$10.8\pm2.84$  &$46.4^{+12.7}_{-12.9}$ &$89.6^{+3.2}_{-2.7}$   &$20.1\pm6.9$
                                        &$23.5^{+5.8}_{-5.9}$  &$3.6^{+1.5}_{-1.3}$  &$38\pm11.4$ \\
\hline
\hline
\end{tabular*}
\end{table}

In Tables \ref{Tab:brexp}-\ref{Tab:other2}, we present our numerical results
for the branching ratios, the direct CP asymmetries, and the polarization
observables of the thirteen $\bar B_s^0\to V V$ decays.
Besides, the dominant contributions to these decays
are also listed in the tables through the symbols ``${\rm T}$" (the color-allowed tree contributions),
``${\rm C}$" (the color-suppressed tree contributions), ``${\rm P}$" (penguin contributions), and ``${\rm A}$" (the annihilation contributions).
The label ``LO" denote the PQCD predictions at the leading order only.  The label  ``NLOWC"
means the LO results with the NLO Wilson coefficients, and $``+{\rm VC}"$, $``+{\rm QL}"$, $``+{\rm MP}"$, and $``{\rm NLO}"$
mean the inclusions of the vertex corrections, the quark loops, the magnetic penguin,
and all the above NLO corrections, respectively. In Table \ref{Tab:ann}, we also test the effects of the contributions from the
annihilation diagrams, and the label ``No Ann"  means the full NLO contributions
except for the annihilation contributions.
For comparison, the experimental measurements~\cite{cdf,lhcbks1,lhcbphiks,lhcbks,lhcbphi,lhcbphi2,lhcbrhophi} and
the numerical results arising from the former PQCD~\cite{pqcd2},
QCDF~\cite{qcdf09,qcdf07}, SCET~\cite{scet}
and Factorization-Assisted Topological-Amplitude Approach(FAT)\cite{fat} are also presented in these tables.
The theoretical errors mainly come from the uncertainties of various input parameters, in particular, the dominant
ones from the shape parameter $\omega_{B_s}=0.50 \pm 0.05$, $f_{B_s}=0.23\pm 0.02$ GeV
and the Gegenbauer moments in the
distribution amplitudes of light vector mesons.
The total errors of the NLO PQCD predictions are given in the Tables
by adding the individual uncertainties in quadrature.

\begin{table}[t]  
\caption{ The PQCD predictions for the relative phase $\phi_{\|}$ (the first row) and $\phi_{\bot}$
(the second row) of the three measured $\bar B_s^0\to \phi \phi, K^{*0} \phi, \bar K^{*0} K^{*0}$ decays.
For  comparison, we also cite the theoretical predictions in the previous PQCD\cite{pqcd2}, SCET\cite{scet} and FAT\cite{fat}
approaches. The experimental data are taken from the Refs. \cite{lhcbphi,lhcbphiks}.}
\label{Tab:phiexp}
\begin{tabular*}{16cm}{@{\extracolsep{\fill}}l|ll|lll|l} \hline\hline
{\rm Mode}&{\rm LO} &{\rm NLO}&{\rm PQCD\cite{pqcd2}}&{\rm SCET\cite{scet}}&{\rm FAT\cite{fat}}&{\rm Data} \\ \hline
$\bar B_s^0\to \phi \phi$&              $2.07$ &$1.65^{+0.20}_{-0.13}$&$2.01\pm0.23 $&$2.41\pm0.62  $&$2.53\pm0.28   $&$2.54\pm0.11$     \\
                                       &$2.10$ &$1.69^{+0.22}_{-0.12}$&$2.00^{+0.24}_{-0.21} $&$2.54\pm0.62  $&$2.56\pm0.27   $&$2.67\pm0.23$   \\

$\bar B_s^0\to K^{*0} \phi$&            $1.98$ &$1.62^{+0.22}_{-0.18}$&$1.95^{+0.21}_{-0.22}  $&$2.37\pm0.59  $&$2.52\pm0.27   $&$1.75\pm0.58$   \\
                                       &$1.98$ &$1.65^{+0.17}_{-0.14}$&$1.95^{+0.21}_{-0.22}  $&$2.50\pm0.59  $&$2.55\pm0.27   $&$$   \\

$\bar B_s^0\to \bar K^{*0} K^{*0}$&     $2.12$ &$1.84^{+0.25}_{-0.20}$&$2.12^{+0.21}_{-0.25}   $&$2.47\pm0.67  $&$2.10\pm0.23   $&$ $\\
                                       &$2.15$ &$1.89^{+0.22}_{-0.21}$&$2.15^{+0.22}_{-0.23}   $&$2.60\pm0.67  $&$2.10\pm0.23  $&$ $   \\
\hline\hline
\end{tabular*}
\end{table}

\begin{table}[t]    
\caption{ The PQCD predictions for the direct CP asymmetries  $\cala_{CP}^{\rm dir}$($\%$)
of $\bar B_s^0\to \phi \phi, K^{*0} \phi, \bar K^{*0} K^{*0}$ decays. }
\label{Tab:dcp01}
\begin{tabular*}{16cm}{@{\extracolsep{\fill}}l|l|ll|llll} \hline\hline
{\rm Mode}&{\rm Class} &{\rm LO}&{\rm NLO}&${\rm PQCD_{LO}}$\cite{pqcd2}&{\rm QCDF}\cite{qcdf09}&{\rm SCET}\cite{scet}&{\rm FAT\cite{fat}} \\ \hline
$\bar B_s^0\to \phi \phi$&         {\rm P}&    $0       $&$0.7\pm0.2              $&$0 $&$0.2^{+0.6}_{-0.3}   $&$-0.39\pm0.44        $&$0.83\pm0.28$\\
$\bar B_s^0\to K^{*0} \phi$&       {\rm P}&    $0       $&$-15.9^{+2.7}_{-2.0}    $&$0 $&$-9^{+5}_{-6}        $&$6.6\pm7.6           $&$-17.3\pm5.6$     \\
$\bar B_s^0\to \bar K^{*0}K^{*0}$& {\rm P}&    $0       $&$0.7\pm0.2              $&$0 $&$0.4^{+0.1}_{-0.6}   $&$-0.56\pm0.61        $&$0.78\pm0.19$     \\
 \hline\hline
\end{tabular*}
\end{table}
\begin{table}[]        
\caption{ The PQCD predictions for CP-averaged branching ratios (in units of $10^{-6}$) of the ten
$\bar B_s^0\to VV$ decays.}
\label{Tab:br}
\begin{tabular*}{16cm}{@{\extracolsep{\fill}}l|l|ll|llll} \hline\hline
{\rm Mode}&{\rm Class} &{\rm LO}&{\rm NLO}&{\rm PQCD}\cite{pqcd2}&{\rm QCDF}\cite{qcdf09}&{\rm SCET}\cite{scet}&{\rm FAT\cite{fat}} \\ \hline
$\bar B_s^0\to K^{*+}\rho^-$&   {\rm T}&     $23.2 $&$20.6^{+9.2}_{-6.9}   $&$24.0^{+11.1}_{-9.4}  $&$21.6^{+1.8}_{-3.5}  $&$28.1\pm 4.2    $&$38.6\pm8.27$    \\
$\bar B_s^0\to K^{*0}\rho^0$&   {\rm C}&     $0.38 $&$0.69^{+0.21}_{-0.16} $&$0.40^{+0.24}_{-0.17} $&$1.3^{+3.2}_{-0.7}   $&$1.04\pm 0.3    $&$1.18\pm0.46$      \\
$\bar B_s^0\to K^{*0}\omega$&   {\rm C}&     $0.33 $&$0.66^{+0.20}_{-0.18} $&$0.35^{+0.19}_{-0.20} $&$1.1^{+2.4}_{-0.6}   $&$0.41\pm 0.14   $&$0.97\pm0.38$    \\
\hline
$\bar B_s^0\to K^{*+}K^{*-}$&   {\rm P}&     $5.02 $&$6.50^{+2.8}_{-2.1}   $&$5.4^{+3.3}_{-2.5}    $&$7.6^{+2.5}_{-2.7}   $&$11.0\pm 3.3    $&$15.9\pm3.5$      \\
$\bar B_s^0\to \omega \phi$&    {\rm P}&     $0.19 $&$0.22^{+0.15}_{-0.10} $&$0.17^{+0.21}_{-0.08} $&$0.18^{+0.13}_{-0.06}$&$0.04\pm 0.01   $&$3.69\pm1.45$      \\
$\bar B_s^0\to \rho^0 \phi $&
                                {\rm P}&     $0.21 $&$0.25^{+0.18}_{-0.11} $&$0.23^{+0.15}_{-0.06} $&$0.18^{+0.8}_{-0.13} $&$0.36\pm 0.05   $&$0.07\pm0.03$      \\
\hline
$\bar B_s^0\to \omega\rho^0$&   {\rm A}&   $0.009$&$0.007                $&$0.009                $&$0.004                  $&$-$   &$0.08\pm0.05$\\
$\bar B_s^0\to \rho^+ \rho^-$&  {\rm A}&   $1.65 $&$1.70^{+0.6}_{-0.5}   $&$1.5^{+0.7}_{-0.6}    $&$0.68^{+0.7}_{-0.5}     $&$-$   &$0.10\pm0.06$\\
$\bar B_s^0\to \rho^0 \rho^0$&  {\rm A}&   $0.82 $&$0.90^{+0.6}_{-0.5}   $&$0.74^{+0.7}_{-0.6}   $&$0.34^{+0.4}_{-0.3}     $&$-$   &$0.05\pm0.03$\\
$\bar B_s^0\to \omega\omega$&   {\rm A}&   $0.45 $&$0.50^{+0.18}_{-0.16} $&$0.40^{+0.21}_{-0.23} $&$0.19^{+0.21}_{-0.15}   $&$-$   &$0.03\pm0.02$\\
 \hline\hline
\end{tabular*}
\end{table}

Among the thirteen $B_s^0 \to VV$ decays considered in this work,
only three of them, namely, $\bar{B}^0_s \to \phi \phi$,
$\bar{B}^0_s \to K^{*0} \phi$ and $\bar{B}^0_s \to \bar K^{*0} K^{*0} $,
have been well measured by experiments up to now. The measured values of the branching ratios, $(f_L,f_\perp)$  and
$(\phi_{\|},\phi_{\bot})$, can be found easily in Table~\ref{Tab:brexp}-\ref{Tab:phiexp}.
For $\bar{B}^0_s \to \rho^0 \phi$ decay, however, only its branching ratio has been reported by LHCb
Collaboration very recently \cite{lhcbrhophi}:
\beq
Br(\bar{B}^0_s \to \rho^0 \phi)=(0.27 \pm 0.08) \times 10^{-6},
\eeq
and other physical parameters are still unknown at present.
On the basis of the data and the theoretical predictions in different approaches/methods, some remarks are in order:
\begin{itemize}
\item[(1)]
Generally speaking, the $\bar B_s^0 \to \phi \phi$ and
$\bar B_s^0  \to \bar K^{*0} K^{*0}$, and the $\bar B_s^0 \to
K^{*0} \phi$  decays are governed
by the QCD penguin contributions through the $b \to s$ and $b \to d$ transition,
respectively. Then the former two CKM-favored modes have larger decay rates
than the latter CKM-suppressed one due to $|V_{ts}/V_{td}|^2 \sim 21$,
which can be easily seen from the Table~\ref{Tab:brexp}.
The evident deviations between the branching ratios of the two $\Delta S=1$ channels,
i.e., $B_s^0 \to \phi \phi$ and $B_s^0 \to K^{*0} \bar{K}^{*0}$, imply the destructive
interferences induced by the significant $SU(3)$ flavor symmetry-breaking effects,
which made the $Br(B_s^0 \to K^{*0} \bar{K}^{*0})$ smaller than the $Br(B_s^0 \to
\phi \phi)$ with a factor about $3$, while the difference of the measured decay rates is around a factor of two
as shown in Table~\ref{Tab:brexp}.

\item[(2)]
From the Table~\ref{Tab:brexp}, one can find that,
the NLO contributions such as Wilson coefficients at NLO level, the
vertex corrections, the quark loop effects can provide the evident
enhancements to the $\bar B_s^0  \to \phi \phi, K^{*0} \phi$ and
$\bar B_s^0  \to \bar K^{*0} K^{*0}$ decays, while the chromo-magnetic penguin
contributions lead to the small reduction to their decay rates.
Furthermore, the quark loop effects largely increase the numerical results of the branching ratios
of the considered three modes because of the possible constructive interferences
between the tree and penguin amplitudes. However, the total enhancements to the branching ratios due to
the inclusion of all known NLO corrections are not very large: less than $35\%$ in magnitude.
Anyway, the consistency between the theory and the data for the decay rates of
the two $\Delta S =1$ modes are improved and the predictions at NLO level agree with the current measurements
within uncertainties. It is worth emphasizing that, for the $\Delta D=1$ $\bar B_s^0  \to  K^{*0} \phi$
decay, the NLO PQCD prediction for its branching fraction is still much
smaller than the present data, however, it agrees well with those
in different theoretical approaches/methods such as QCDF, SCET, and FAT within errors,
which can be seen explicitly in Table \ref{Tab:brexp}. It is expected that
the combined analyses from the updated LHCb and Belle-II measurements in the
near future would help to clarify this discrepancy.

\item[(3)]
For $B_s^0 \to \bar{K}^{*0} K^{*0}$ decay, the PQCD predictions for $f_L$ and $f_\perp$  will become
a bit large (small) after the inclusion of the NLO contributions, but still be consistent
with previous theoretical predictions based on QCDF, SCET and FAT, even with those measured ones,
since both theoretical and experimental errors are still rather large in magnitude.
For $B_s^0 \to \phi\phi$ and $B_s^0 \to \phi \bar{K}^{*0}$ decays, fortunately, it is more interesting to
observe that the NLO contributions to $f_L$ and $f_\perp$  are very small in size,
while the PQCD predictions for both $f_L$ and $f_\perp$ agree well with other theoretical predictions,
and with those measured values as well.

\item[(4)]
As we know, the annihilation diagrams can play important roles in the investigations of heavy flavor system, although these contributions are generally power suppressed. As mentioned in the Introduction, the weak penguin annihilation contributions can be considered as one of the strategies to explain the ``polarization puzzle". In fact,
when the annihilation amplitudes are turned off in the decays of $\bar B_s^0  \to \phi \phi, K^{*0} \phi$ and $\bar B_s^0  \to \bar K^{*0} K^{*0}$, all the branching ratios
will decrease about $60\%$, $64\%$, and $27\%$, respectively, which could be easily inferred from the Table \ref{Tab:ann}. Correspondingly, without the annihilation
contributions, the longitudinal-polarization-dominance really exhibits, which suggests that the annihilation contributions in these penguin-dominated $B_s$ decay modes could indeed enhance the transverse polarization fractions and reduce the longitudinal ones simultaneously with different extent. Of course, more stringent constraints on the theoretical uncertainties arising from the nonperturbative hadronic parameters are urgently demanded.
Although the predictions look roughly consistent with the current measurements within
still large theoretical errors, it should be noted that the annihilation amplitudes might not be the only source to explain the dramatically small $f_L(B_s^0 \to \bar{K}^0 K^{*0})$, if the significantly large differences between the theory and the experiment always exists as given in the Table \ref{Tab:fexp}.

\item[(5)]
Moreover, the direct {\it CP} asymmetries and the relative phases of the
decays of $\bar B_s^0  \to \phi \phi, K^{*0} \phi$  and
$\bar B_s^0  \to \bar K^{*0} K^{*0}$ are also studied in the PQCD approach with
inclusion of the currently known NLO contributions.
Because these considered modes are induced only by penguin operators,
their direct {\it CP}-violations are naturally zero without the interferences between
the tree and penguin amplitudes in PQCD approach at LO, as listed in Table \ref{Tab:dcp01}.
After the inclusion of the NLO contributions, their direct CP asymmetries are nonzero but still very small:
$(0.7\pm 0.2)\%$, $(-15.9^{+2.7}_{-2.0})\%$, and $(0.7 \pm 0.2)\%$ respectively, which are comparable
with the results of QCDF~\cite{qcdf09}
($(0.2^{+0.6}_{-0.4})\%$, $(-9^{+5}_{-6})\%$, and $(0.4^{+1.0}_{-0.6})\%$)
and SCET~\cite{scet}($(-0.39\pm 0.44)\%$, $(6.6 \pm 7.6)\%$, and $(-0.56\pm 0.61)\%$)
but with an overall opposite sign to those in SCET. In light of the relative phases,
the NLO PQCD predictions of $\phi_\parallel$ and $\phi_\perp$ of the $B_s^0 \to \phi \phi$ mode are
a bit smaller than the measured one, which would be further studied in the future.
The numerical results for other relative phases would be tested by the near future experiments.
\end{itemize}

\begin{table}[t]  
\caption{ The PQCD predictions for the longitudinal polarization fractions $f_L$($\%$)
of the remaining ten $\bar B_s^0 \to VV $ decays.}
\label{Tab:fl}
\begin{tabular*}{16cm}{@{\extracolsep{\fill}}l|l|ll|llll} \hline\hline
{\rm Mode}&{\rm Class} &{\rm LO}&{\rm NLO}&{\rm PQCD}\cite{pqcd2}&{\rm QCDF}\cite{qcdf09}&{\rm SCET}\cite{scet}&{\rm FAT\cite{fat}} \\ \hline
$\bar B_s^0\to K^{*+}\rho^-$&      {\rm T}&      $93.4   $&$94.1^{+1.0}_{-1.0}   $&$95^{+1.4}_{-1.4}    $&$92^{+1.4}_{-3.6}    $&$99.1\pm 0.3   $&$94.4\pm1.2$  \\
$\bar B_s^0\to K^{*0}\rho^0$&      {\rm C}&      $50.1   $&$83.4^{+4.8}_{-4.8}   $&$57^{+8.1}_{-15.7}   $&$90^{+5.0}_{-24.0}   $&$87\pm 5       $&$79.8\pm8.0$\\
$\bar B_s^0\to K^{*0}\omega$&      {\rm C}&      $51.7   $&$82.7^{+5.4}_{-5.3}   $&$50^{+13.1}_{-17.0}  $&$90^{+4.2}_{-23.2}   $&$64\pm 15      $&$77.9\pm9.2$\\
\hline
$\bar B_s^0\to K^{*+}K^{*-}$&      {\rm P}&      $40.2   $&$48.1^{+9.7}_{-8.9}   $&$42^{+14.2}_{-11.2}  $&$52^{+20.2}_{-21.6}  $&$55\pm 14      $&$30.9\pm10.4$\\
$\bar B_s^0\to \omega \phi$&       {\rm P}&      $65.7   $&$55.2^{+6.6}_{-4.7}   $&$69^{+11.2}_{-12.6}  $&$95^{+1.0}_{-42.1}   $&$100$&$-$\\
$\bar B_s^0\to \rho^0 \phi$&       {\rm P}&      $84.5   $&$90.2^{+1.2}_{-1.5}   $&$86^{+1.4}_{-1.4}    $&$88^{+2.2}_{-18.0}   $&$100$&$-$\\
\hline
$\bar B_s^0\to \rho^+ \rho^-$&     {\rm A}&    $\sim 100  $&$\sim 100    $&$\sim 100    $&$\sim 100     $&$-$&$-$\\
$\bar B_s^0\to \rho^0 \rho^0$&     {\rm A}&    $\sim 100  $&$\sim 100    $&$\sim 100    $&$\sim 100     $&$-$&$-$\\
$\bar B_s^0\to \omega\omega$&      {\rm A}&    $\sim 100  $&$\sim 100    $&$\sim 100    $&$\sim 100     $&$-$&$-$\\
$\bar B_s^0\to \omega\rho^0$&      {\rm A}&    $\sim 100  $&$\sim 100    $&$\sim 100    $&$\sim 100     $&$-$&$-$\\
 \hline\hline
\end{tabular*} \end{table}
\begin{table}[b]    
\caption{ The PQCD predictions for the direct CP asymmetries  $\cala_{CP}^{\rm dir}$($\%$)
of the all thirteen $\bar B_s^0 \to VV $ decays. }
\label{Tab:dcp}
\begin{tabular*}{16cm}{@{\extracolsep{\fill}}l|l|ll|llll} \hline\hline
{\rm Mode}&{\rm Class} &{\rm LO}&{\rm NLO}&${\rm PQCD_{LO}}$\cite{pqcd2}&{\rm QCDF}\cite{qcdf09}&{\rm SCET}\cite{scet}&{\rm FAT\cite{fat}} \\ \hline
$\bar B_s^0\to K^{*+}\rho^-$&      {\rm T}&    $-8.5    $&$-13.7^{+3.3}_{-2.8}    $&$-9.1^{+1.7}_{-2.6}       $&$-11^{+4.1}_{-1.4}   $&$-7.7\pm 9.2   $&$-10.9\pm3.0$\\
$\bar B_s^0\to K^{*0}\rho^0$&      {\rm C}&    $65.4    $&$59.1^{+10.4}_{-8.9}    $&$62.7^{+15.7}_{-20.4}     $&$46^{+23.4}_{-39.2}  $&$19.5\pm 23.5  $&$4.9\pm18.3$\\
$\bar B_s^0\to K^{*0}\omega$&      {\rm C}&    $-73.3   $&$-69.3^{+11.3}_{-10.6}  $&$-78.1^{+17.6}_{-14.1}    $&$-50^{+28.1}_{-17.2} $&$-36.8\pm 40.1 $&$32.2\pm16.0$\\
\hline
$\bar B_s^0\to K^{*+}K^{*-}$&      {\rm P}&    $4.5     $&$6.8^{+5.4}_{-4.3}      $&$8.8^{+2.6}_{-9.8}        $&$21^{+2.2}_{-4.5}    $&$20.6\pm 23.3  $&$21.1\pm7.1$\\
$\bar B_s^0\to \omega \phi$&       {\rm P}&    $23.5    $&$-22.1^{+1.0}_{-1.0}    $&$28.0^{+4.9}_{-8.2}       $&$-8^{+20.2}_{-15.1}  $&$0     $&$-15.0\pm7.0$    \\
$\bar B_s^0\to \rho^0 \phi$&       {\rm P}&    $18.7    $&$39.6^{+3.6}_{-3.1}     $&$-4.3^{+1.9}_{-1.3}       $&$83^{+10.1}_{-3.6}   $&$0     $&$0$\\
\hline
$\bar B_s^0\to \rho^+ \rho^-$&     {\rm A}&            $-2.1    $&$-0.4^{+0.5}_{-0.3}     $&$-2.9^{+1.6}_{-1.7}     $&$0     $&$-$&$0$\\
$\bar B_s^0\to \rho^0 \rho^0$&     {\rm A}&            $-2.1    $&$-0.4^{+0.5}_{-0.3}     $&$-2.9^{+1.6}_{-1.7}     $&$0     $&$-$&$0$\\
$\bar B_s^0\to \omega\omega$&      {\rm A}&            $-1.9    $&$-0.4^{+0.5}_{-0.3}     $&$-3.3^{+1.8}_{-1.7}     $&$0     $&$-$&$0$\\
$\bar B_s^0\to \omega\rho^0$&      {\rm A}&            $7.3     $&$5.8^{+3.2}_{-3.4}      $&$11.1^{+2.7}_{-6.6}     $&$0     $&$-$&$0$\\
 \hline\hline
\end{tabular*}
\end{table}

We now turn to study the remaining ten  $B_s \to VV$ decays that have not been measured experimentally.
In Table~\ref{Tab:br}, we present our LO and NLO PQCD predictions for the CP-averaged branching ratios of
the ten $\bar B_s^0\to V V$ decay modes. We also classify these modes with different dominant topologies
such as ``$T$", ``$C$", ``$A$", etc.. We find numerically that:
\begin{itemize}
\item[(1)]
Explicitly, the ten $B_s^0 \to VV$ decays as listed in Table~\ref{Tab:br} can be classified into three types:
(a) the one ``T" decay $\bar B_s^0 \to \rho^- K^{*+}$ and two ``C" decays $\bar B_s^0 \to (\rho^0, \omega) K^{*0}$;
(b) three ``P" decays $\bar B_s^0 \to K^{*+} K^{*-}$ and $\bar B_s^0 \to (\rho^0, \omega) \phi$ modes; and
(c) four pure weak annihilation ``A" decays $\bar B_s^0 \to (\rho, \omega) (\rho, \omega)$.
In fact, we here have reproduced the predictions of the branching ratios as given in
Ref.~\cite{pqcd2} with PQCD approach at LO independently.
The slightly small deviations appeared in the Table~\ref{Tab:br}
are induced by some updated input parameters, such as the decay constants
and the CKM matrix elements.

\item[(2)]
For  $\bar B_s^0 \to \rho^- K^{*+}$ decay, its LO decay rate
will decrease around $10\%$ after inclusion of the known NLO corrections
due to less sensitivity to the vertex corrections.
Hence, the NLO PQCD prediction of $Br(\bar B_s^0 \to \rho^- K^{*+})$  is still
consistent with those in the QCDF, SCET, even FAT approaches within the theoretical uncertainties.
However, because the ``C" channels are highly sensitive
to the vertex corrections with a large imaginary amplitude to the factorizable emission diagrams,
the $\bar B_s^0 \to \rho^0 K^{*0}$ and $\omega K^{*0}$ decay rates
receive significant enhancements with a factor around 2, which can be seen
clearly in the Table~\ref{Tab:br} and agree with those estimated in other
approaches such as QCDF, SCET, and FAT in general.
Moreover, the relation of $Br(\bar B_s^0 \to \rho^0 K^{*0}) \sim Br(\bar B_s^0 \to \omega K^{*0})$
is induced by adopting the same QCD behavior of the $\rho^0$ and $\omega$ states
and similar decay constants and meson masses, which can be
observed from the input parameters in Eqs.~(\ref{eq:Geb}) and~(\ref{eq:para}).

\item[(3)]
For the decays of $\bar B_s^0\to \omega \phi, \rho^0 \phi$ and $\bar B_s^0\to K^{*+}K^{*-}$,
only the measured decay rate $Br(\bar B_s^0 \to \rho^0 \phi)=(0.27\pm 0.08)\times 10^{-6}$ from LHCb
\cite{lhcbrhophi} is available now.
From Table~\ref{Tab:br}, one can see easily that the predicted branching ratio
in the PQCD approach at the LO and NLO level agrees well with the data.
Of course, all the available predictions for the $Br(B_s^0 \to \rho^0 \phi)$
in the framework of the QCD-based factorization approaches also be consistent with
the experimental measurements within the errors.
However, the prediction based on the FAT is much smaller.
It is more interesting to find that different patterns between
$Br(B_s^0 \to \rho^0 \phi)$ and $Br(B_s^0 \to \omega \phi)$ have been predicted
in various frameworks: the moderate interferences between the $\bar u u$ and $\bar d d$
components in the $\rho$ and $\omega$ mesons result in the relation of $Br(B_s^0 \to
\rho^0 \phi) \sim Br(B_s^0 \to \omega \phi)$
in PQCD$_{\rm LO}$, PQCD$_{\rm NLO}$, and QCDF, respectively, but the strong effects
with different destructive and/or constructive interferences lead to the relations
of $Br(B_s^0 \to \rho^0 \phi)_{\rm SCET} \gg Br(B_s^0 \to \omega \phi)_{\rm SCET}$
and $Br(B_s^0 \to \rho^0 \phi)_{\rm FAT} \ll Br(B_s^0 \to \omega \phi)_{\rm FAT}$.
Moreover, the improved NLO PQCD prediction of $Br(B_s^0 \to K^{*+} K^{*-})$ is also
consistent with that provided by other approaches/methods.
These phenomenologies would be tested in the near future at the LHCb and Belle-II
experiments by measuring the $Br(B_s^0 \to \omega \phi)$ with good precision.

\item[(4)]
For the four pure annihilation decays $\bar B_s^0\to (\rho^+\rho^-, \rho^0\rho^0,\rho^0\omega,\omega\omega)
$, in fact, the NLO correction comes only from the usage of
the NLO Wilson coefficients $C_i(\mu)$ and the strong coupling constant
$\alpha_s(\mu)$ at the two-loop level, which result in negligible corrections to the
$Br(B_s^0 \to \rho^0 \omega)$ and $Br(B_s^0 \to \rho^+ \rho^-)$ while around $10\%$
enhancement to the $Br(B_s^0 \to \rho^0 \rho^0)$ and $Br(B_s^0 \to \omega \omega)$.
It should be mentioned that the annihilation diagrams
in the QCDF and SCET framework have to be fitted from the experimental measurements
because of the endpoint singularities. While, in Ref.~\cite{scet}, the authors
neglected the contributions arising from the annihilation diagrams based on the
arguments of the ${\cal O}(1/m_B)$ power-suppressed effects. The very different
results in the FAT method from those in the PQCD and QCDF approaches should
be examined by the near future measurements at LHC and Belle-II experiments.
It is worth emphasizing that the pure annihilation $B_s^0 \to \pi^+ \pi^-$ decay rate
has been confirmed by the CDF and LHCb collaborations. Therefore, these large
branching ratios of the $B_s^0 \to \rho^+ \rho^-$, $\rho^0 \rho^0$, and $\omega \omega$ modes are
expected to be verified soon. Moreover, the substantial cancelations between
the contributions arising from the $\bar u u$ and $\bar d d$ components of the $\rho^0$ and $\omega$
mesons result in the tiny decay rate of the $B_s^0 \to \rho^0 \omega$ mode, which would be examined in the future.

\end{itemize}

Next, we turn to discuss the longitudinal polarization fractions
$f_L$ of the remaining ten $B_s \to VV$ decays.
From the numerical results as given in Table~\ref{Tab:fl}, one can see that:
\begin{itemize}
\item[(1)]
Generally speaking, except for the $B_s^0 \to K^{*+} K^{*-}$ and $B_s^0 \to \omega \phi$ channels, most of these considered ten $B_s^0 \to VV$ decays are governed
by the longitudinal amplitudes by including the known NLO corrections in the PQCD
approach, in which (a) the $B_s^0 \to K^{*0} (\rho^0, \omega)$ decays do receive
significant enhancements to $f_L$, then both of the fractions
are increased from around $50\%$ to about $80\%$ and consistent with those predicted
in the QCDF, SCET, and FAT within errors; (b) the four pure annihilation $B_s^0 \to
(\rho^+\rho^-,\rho^0\rho^0, \rho^0\omega,\omega \omega)$ decays are absolutely dominated by the longitudinal
polarization contributions, therefore, the fractions of these four modes are
around $100\%$.

\item[(2)]
For the penguin-dominated $\bar B_s^0\to K^{*+} K^{*-}$ decay, we find a small LO PQCD prediction
$f_L \perp \sim 0.4$, as presented in the Table~\ref{Tab:fl}.
When the NLO contributions are taken into account, $f_L$ will become a little larger to around $0.48$.
It is interesting to note that the
significant transverse-components dominance have been obtained in various
approaches such as QCDF, SCET, and FAT, which could be examined in the
near future by experiments associated with the large decay rates predicted in the
aforementioned approaches.

\item[(3)]
For the pure emission decay of $\bar B_s^0\to \omega \phi$, the longitudinal polarization fraction
$f_L$ is around $55 \%$ in the PQCD approach at NLO level,
because the $(S-P)(S+P)$ densities in the hard spectator scattering diagrams
together with the NLO contributions can provide the sizable transverse
polarization contributions. By considering the vary large theoretical errors,
the relation $f_L \sim (f_{||}+f_{\perp})$ might be got in the framework
of QCDF. However, it is highly different from that provided in SCET, namely,
$100 \%$. The future stringent tests from the experimental measurements would
help us to distinguish these theoretical approaches.
Of course, it seems not easy
because of the predicted small branching ratios around $10^{-7} \sim 10^{-8}$ in various approaches.
\end{itemize}

\begin{table}[t]      
\caption{Transverse polarization fractions $f_{\bot}$($\%$) , and
relative phase $\phi_{\|}$(\rm rad) and $\phi_{\bot}$(\rm rad)
in the $\bar B_s^0 \to VV $ decays calculated in the PQCD approach.}
\label{Tab:other1}
\begin{tabular*}{14cm}{@{\extracolsep{\fill}}l|ll|ll|ll}
\hline\hline
 &
 \multicolumn{2}{c|}{$f_\bot$(\%)} & \multicolumn{2}{c|}{$\phi_\|$(\rm rad)}&
 \multicolumn{2}{c}{$\phi_\bot$(\rm rad)}\\
 \hline
 {Decay Mode}&{LO}&{NLO}&{LO}&{NLO}&{LO}&{NLO}\\
 \hline
 $\bar B_s^0\to  \rho^- K^{*+}$           &$3.2$   &$2.8^{+0.23}_{-0.45}$  &$3.21$    &$3.12^{+0.08}_{-0.06}$  &$3.21$   &$3.18^{+0.03}_{-0.05}$     \\
 $\bar B_s^0\to  \rho^0 K^{*0}$           &$26.2$   &$8.9^{+2.1}_{-1.9}$ &$1.63$    &$1.50^{+0.64}_{-0.12}$  &$1.63$   &$1.54^{+0.41}_{-0.11}$      \\
 $\bar B_s^0\to  \omega K^{*0}$           &$25.4$   &$8.9^{+0.05}_{-0.05}$     &$2.38$   &$1.85^{+0.05}_{-0.05}$  &$2.45$   &$1.94^{+0.05}_{-0.05}$       \\
\hline
 $\bar B_s^0\to  K^{*+} K^{*-}$           &$27.3$   &$23.9^{+4.4}_{-5.2}$     &$2.79$   &$2.34^{+0.13}_{-0.10}$  &$2.83$   &$2.39^{+0.30}_{-0.22}$        \\
 $\bar B_s^0\to  \omega \phi $            &$17.3$    &$38.6^{+5.6}_{-3.2}$     &$2.93$   &$3.12^{+0.23}_{-0.12}$  &$2.92$   &$3.10^{+0.25}_{-0.13}$       \\
 $\bar B_s^0\to  \rho^0 \phi $            &$7.8$    &$4.4^{+1.5}_{-1.1}$ &$2.52$   &$2.41^{+0.07}_{-0.05}$  &$2.50$   &$2.67^{+0.05}_{-0.04}$      \\
 \hline
 $\bar B_s^0\to  \rho^+ \rho^- $          &$\sim 0$    &$\sim 0$     &$3.42$   &$3.35^{+0.21}_{-0.16}$  &$3.02$   &$2.67^{+0.15}_{-0.20}$    \\
 $\bar B_s^0\to  \rho^0 \rho^0 $          &$\sim 0$    &$\sim 0$     &$3.42$   &$3.35^{+0.21}_{-0.16}$  &$3.02$   &$2.67^{+0.15}_{-0.20}$    \\
 $\bar B_s^0\to  \rho^0 \omega $          &$\sim 0$    &$\sim 0$     &$3.38$   &$3.28^{+0.05}_{-0.03}$  &$2.34$   &$2.02^{+0.15}_{-0.14}$    \\
 $\bar B_s^0\to  \omega \omega $          &$\sim 0$    &$\sim 0$     &$3.41$   &$3.13^{+0.18}_{-0.24}$  &$3.05$   &$2.66^{+0.16}_{-0.13}$   \\
\hline\hline
\end{tabular*}
\end{table}

\begin{table}[b]   
\caption{The relative phases $\Delta\phi_{\parallel}$($10^{-2}$ \rm rad),
$\Delta\phi_{\bot}$($10^{-2}$ \rm rad), and the CP asymmetry parameters
$A^0_{CP}$($\%$) and $A^\bot_{CP}$($\%$) in the $\bar B_s^0 \to VV $ decays calculated in the LO and NLO PQCD approach.}
\label{Tab:other2}
\begin{tabular*}{16cm}{@{\extracolsep{\fill}}l|ll|ll|ll|ll}
\hline \hline
 &
 \multicolumn{2}{c|}{$A^0_{CP}$(\%)}&
 \multicolumn{2}{c|}{$A^\bot_{CP}$(\%)}&
 \multicolumn{2}{c|}{$\Delta\phi_{\parallel}(10^{-2} \rm rad)$}&
 \multicolumn{2}{c}{$\Delta\phi_{\bot}(10^{-2} \rm rad)$}\\
 \hline
 {Decay Mode}&{LO}&{NLO}&{LO}&{NLO}&{LO}&{NLO}&{LO}&{NLO}\\  \hline
 $\bar B_s^0\to \phi \phi$                 &$0$  &$0.5^{+0.5}_{-0.3}$  &$0$
                                         &$-0.3^{+0.2}_{-0.6}$    &$0$        &$\sim 0$   &$0$     &$\sim 0$   \\
 $\bar B_s^0\to K^{*0} \phi$                &$0$
                                          &$-6.3^{+1.8}_{-1.4}$ & $0$&$5.8^{+2.1}_{-2.4}$   &$\sim 0$&$-5.4^{+1.9}_{-1.4}$ &$\sim 0$   &$-4.8^{+2.9}_{-2.4}$              \\
 $\bar B_s^0\to \bar K^{*0} K^{*0}$          &$0$  &$0.3^{+0.2}_{-0.2}$
                                        &$0$  &$-0.2^{+0.3}_{-0.2}$&$0$      &$\sim 0$   &$0$       &$\sim 0$     \\
 \hline
 $\bar B_s^0\to  \rho^- K^{*+}$              &$-3.6$
                                          &$-4.4^{+0.6}_{-0.7}$   &$50.1$   &$64.6^{+8.2}_{-8.3} $&$11.3$   &$15.5^{+3.7}_{-3.3}$  &$10.0$     &$13.5^{+4.3}_{-2.4}$     \\
 $\bar B_s^0\to  \rho^0 K^{*0}$              &$-18.3$
                                          &$-11.2^{+6.1}_{-5.7}$ &$20.3$  &$35.3^{+7.2}_{-9.5}$&$-31.1$  &$-20.6^{+45.3}_{-11.1}$ &$-35.4$   &$-22.3^{+54.1}_{-13.2}$     \\
 $\bar B_s^0\to  \omega K^{*0}$             &$-9.6$
                                          &$-7.7^{+5.3}_{-6.3}$   &$10.8$  &$28.2^{+7.2}_{-5.3}$&$24.5$   &$29.3^{+9.3}_{-8.9}$     &$28.3$   &$15.5^{+11.6}_{-6.7}$     \\
\hline
 $\bar B_s^0\to  K^{*+} K^{*-}$             &$37.1$
                                          &$34.8^{+17.2}_{-16.3}$   &$-28.2$  &$-23.4^{+2.4}_{-2.2}$ &$65.5$   &$44.2^{+6.1}_{-5.8}$     &$65.7$   &$44.2^{+5.1}_{-4.2}$     \\
$\bar B_s^0\to  \omega \phi $               &$-2.3$
                                          &$-12.8^{+2.6}_{-4.3}$   &$5.9$  &$20.1^{+3.4}_{-2.1}$ &$-33.5$   &$-31.5^{+9.5}_{-11.3}$&$-34.5$   &$-31.5^{+12.3}_{-12.6}$     \\
 $\bar B_s^0\to  \rho^0 \phi $            &$4.5$  &$7.6^{+3.6}_{-3.1}$
                                          &$-25.6$  &$-35.4^{+6.8}_{-6.4}$&$-52.1$   &$-44.5^{+7.2}_{-6.7}$     &$-55.1$   &$-42.3^{+11.2}_{-8.8}$     \\
 \hline
 $\bar B_s^0\to  \rho^+ \rho^- $            &$0.0$  &$0.0$   &$25.2$
                                          &$35.5^{+7.9}_{-5.1}$&$2.8$       &$1.8^{+0.4}_{-0.8}$       &$-27.1$   &$-44.1^{+4.5}_{-3.6}$     \\
 $\bar B_s^0\to  \rho^0 \rho^0 $            &$0.0$  &$0.0$   &$25.2$
                                          &$35.5^{+7.9}_{-5.1}$&$2.8$       &$1.8^{+0.4}_{-0.8}$       &$-27.1$   &$-44.1^{+4.5}_{-3.6}$     \\
 $\bar B_s^0\to  \rho^0 \omega $           &$0.0$  &$0.0$   &$33.3$
                                          &$21.6^{+5.8}_{-4.6}$ &$-10.0$     &$-12.5^{+3.3}_{-3.8}$      &$-28.3$  &$-23.4^{+7.1}_{-8.1}$     \\
 $\bar B_s^0\to  \omega \omega $             &$0.0$  &$0.0$   &$23.9$
                                         &$36.2^{+6.8}_{-4.6}$ &$2.5$     &$1.8^{+0.5}_{-0.5}$      &$-30.5$  &$-44.5^{+3.9}_{-4.4}$     \\
\hline\hline
\end{tabular*}
\end{table}

For the direct CP asymmetries of the  the considered $\bar B_s^0 \to VV$ decays
collected in Table~\ref{Tab:dcp01} and ~\ref{Tab:dcp}, we have some comments as follows:
\begin{itemize}
\item[(1)]
In fact, the LO PQCD predictions for the direct {\it CP} asymmetries of the decays of $B_s^0 \to VV$
obtained in this paper do agree very well with those as given in Ref.~\cite{pqcd2},
except for the $\bar B_s^0\to \rho^0 \phi$ channel.
Due to the different choices of the updated input parameters, the sensitivity of the
direct {\it CP} violation to the adopted parameters can be observed in the
$B_s^0 \to \rho^0 \phi$ mode, and finally the result has an opposite sign to that
in the previous LO PQCD calculations, which demands the tests from the experiments at LHCb and Belle-II.

\item[(2)]
Generally speaking, except for the penguin-dominated $B_s^0 \to \rho^0 \phi, \omega \phi$
and $B_s^0 \to K^{*0} \phi$ modes, the effects of the NLO contributions to the direct {\it CP}
asymmetries are not significant in magnitude for most of the $\bar B_s^0\to VV$ decays.
Specifically, for the $\bar B_s^0\to \omega \phi$ decay,
the PQCD prediction of the $\cala_{CP}^{\rm dir}$ can vary from $20\%$ to $-20\%$,
after the inclusion of the NLO corrections, which is because an extra strong phase
appears in the decay amplitudes from the factorizable emission diagrams.
For the $\bar B_s^0\to \rho^0 \phi$ decay, on the other hand,
the NLO contributions to the $\cala_{CP}^{\rm dir}$ in magnitude can be found
with a factor around 2, relative to the LO PQCD  prediction, which indicates
a possibly constructive interference between the tree and penguin amplitudes after
the inclusion of the NLO vertex corrections.
All the predictions of the $A_{\it CP}^{\rm dir}(B_s^0 \to K^{*0} \phi)$ in various approaches
are generally consistent within large theoretical uncertainties, which could be tested by
the future measurements.

\item[(3)]
For the two "Color-suppressed" decays $\bar B_s^0\to K^{*0}\rho^0$ and $\bar B_s^0\to K^{*0}\omega$ ,
because of the large penguin contributions from the chirally enhanced annihilation diagrams, which are
at the same level as the tree contributions from the emission diagrams, this sizable interference between the
tree and penguin contributions makes the direct CP asymmetries as large as
$60\% \sim 70\%$ but with an opposite sign for these two channels.

\item[(4)]
By comparing the numerical results as listed in the fifth to seventh columns of Table~\ref{Tab:dcp},
due to the different origins of the strong phase,
one can see that the PQCD, QCDF, and SCET predictions for the CP asymmetries of the considered decays are indeed quite
different. As is well known, besides the weak CKM phases, the direct CP asymmetries
also depend on the strong phase.
In SCET, the strong phase is only from the nonperturbative charming penguin at leading power and leading order;
while in the QCDF and PQCD approach, the strong phase comes from the hard spectator scattering and
annihilation diagrams respectively.
The forthcoming LHCb and Belle-II measurements for these direct {\it CP} violations  can
help us to differentiate these factorization approaches.

\item[(5)]
The direct {\it CP}-violating asymmetries, the relative phases, and the differences of the relative phases in different polarizations for the considered $B_s^0 \to VV$
decays have not been measured experimentally to date yet, neither reported in other approaches by the colleagues theoretically. All these predictions in the PQCD approach at NLO level have to await for the future confirmations arising from both of the theoretical and experimental sides.

\end{itemize}

In Table ~\ref{Tab:other1} and~\ref{Tab:other2}, we listed the LO and NLO PQCD predictions for the
transverse polarization fractions $f_{\bot}$, the relative phase $\phi_{\|}$ and $\Delta\phi_{\parallel}$,
$\phi_{\bot}$  and $\Delta\phi_{\bot}$, and the CP asymmetry parameters $A^0_{CP}$($\%$) and $A^\bot_{CP}$($\%$),
for the considered $\bar B_s^0 \to VV $ decays. It is easy to see
that the NLO contributions to all these physics parameters
are small or moderate in magnitude.
All these PQCD predictions could be tested in the near future by the
forthcoming LHCb and Belle-II experiments.

\section{SUMMARY}\label{sec:4}

In this work, we studied the tao-body  charmless hadronic decays $\bar B_s^0\to V V$ ( here $V=(\rho, K^*, \phi,\omega)$)
by employing the PQCD factorization approach with the inclusion of all currently known NLO contributions, such as the
NLO vertex corrections, the quark loop effects and the chromo-magnetic penguin diagrams etc.
We focus on the examination for the effects of those NLO contributions to the {\it CP}-averaged
branching ratios, the {\it CP}-violating asymmetries, the polarization fractions and other physical
observables of the thirteen $\bar B_s^0\to V V$ decay modes.

By the numerical evaluations and the phenomenological analyses, we found the following interesting points:

\begin{enumerate}

\item[(1)]
For the measured $B_s^0 \to \phi\phi, K^{*0} \phi$ and $\bar{K}^{*0} K^{*0}$ decays,
the agreement between the PQCD predictions for the CP-averaged branching ratios
and the measured vales are improved effectively
after the inclusion of the NLO contributions.
For $B_s^0 \to K^{*0} \phi$ decay, although there exists a clear difference
between the central value
the NLO PQCD prediction for its CP-averaged branching ratio
($(0.42^{+0.13}_{-0.10}) \times 10^{-6}$)
and the measured one ( $(1.13\pm 0.30) \times 10^{-6}$), but they are still consistent
within $3\sigma$, due to the still large experimental errors.

\item[(2)]
For the measured $B_s^0 \to \phi\phi, K^{*0} \phi$ and $\bar{K}^{*0} K^{*0}$ decays,
the NLO corrections to the PQCD predictions for the
longitudinal and transverse polarization fractions
$(f_L,f_\perp)$, the relative phases $(\phi_{\|},\phi_{\bot})$ are small in size.
The NLO PQCD
predictions for these physical observables do agree with those from the QCDF, SCET
and FAT approaches, and also agree well with those currently available experimental
measurements.
It ie easy to see from the results as listed in Table III that the weak penguin
annihilation contributions play an important role in understanding the data about
the decay rates, $f_L$ and $f_\perp$ for three measured decays.

\item[(3)]
For $B_s^0 \to \rho^0 \phi$ decay, furthermore,  the NLO PQCD prediction for its
branching ratio  does agree very well with the measured one as reported
by LHCb Collaboration very recently \cite{lhcbrhophi}.

\item[(4)]
For other considered $B_s^0 \to VV$ decays,  the NLO PQCD predictions for the decay rates and
other  physical observables studied in this paper are also
basically consistent with other theoretical predictions obtained based on
QCDF, SCET and FAT approaches/methods.
The future measurements with good precision could be
employed to test or examine the differences among these rather different approaches.
Of course, the still missing NLO
contributions in the PQCD approach are the urgent meanwhile challenging works to be
completed.

\end{enumerate}

\begin{acknowledgments}

This work is supported by the National Natural Science
Foundation of China under Grants  No.~11775117, 11765012 and 11235005,
by the Qing Lan Project of Jiangsu Province for outstanding teachers, and by the Research
Fund of Jiangsu Normal University under Grant No.~HB2016004.

\end{acknowledgments}



\end{document}